\documentclass{article}

\usepackage{arxiv}

\usepackage[utf8]{inputenc} 
\usepackage[T1]{fontenc}    
\usepackage{hyperref}       
\usepackage{url}            
\usepackage{booktabs}       
\usepackage[super,comma,compress]{natbib}
\usepackage{longtable}
\usepackage{rotating}
\usepackage{placeins}
\usepackage{wrapfig}
\usepackage{multirow}
\usepackage{textgreek}
\usepackage{pdfpages}
\usepackage{siunitx} 
\usepackage{comment}
\usepackage{mhchem}
\usepackage{subcaption}

\title{Unsupervised hyperspectral data mining and bioimaging by information entropy and self-modeling curve resolution}

\author{Simon Vilms Pedersen\thanks{Current address: Department of Materials, Imperial College London, UK.} \\
	Department of Green Technology\\
    University of Southern Denmark\\
	Odense, Denmark \\
	\And
	Anders R. Walther \\
	Department of Green Technology\\
    University of Southern Denmark\\
	Odense, Denmark \\
	\AND
	Anthony Callanan\thanks{Current address: Institute for Bioengineering, University of Edinburgh, Scotland.} \\
	Department of Materials and Department of Bioegnineering\\
    Imperial College London\\
    London, UK
    \AND
	Molly M. Stevens \\
	Department of Materials and Department of Bioegnineering\\
    Imperial College London\\
    London, UK
    \AND
	Martin A.B. Hedegaard \\
	Department of Green Technology\\
    University of Southern Denmark\\
	Odense, Denmark \\
    \AND
	Eva C. Arnspang \\
	Department of Green Technology\\
    University of Southern Denmark\\
	Odense, Denmark \\
    Email: \texttt{arnspang@igt.sdu.dk} \\
}

\renewcommand{\shorttitle}{}

\hypersetup{
pdftitle={Unsupervised hyperspectral data mining and bioimaging by information entropy and self-modeling curve resolution},
pdfsubject={stat},
pdfauthor={Simon Vilms Pedersen, Anders R. Walther, Anthony Callanan, Molly M. Stevens, Martin A.B. Hedegaard, Eva C. Arnspang},
pdfkeywords={Dimensionality, spectral unmixing, hyperspectral microscopy, imaging, endmember extraction},
}

\begin{document}
\maketitle

\begin{abstract}
Unsupervised estimation of the dimensionality of hyperspectral microspectroscopy datasets containing pure and mixed spectral features, and extraction of their representative endmember spectra, remains a challenge in biochemical data mining. We report a new versatile algorithm building on semi-nonnegativity constrained self-modeling curve resolution and information entropy, to estimate the quantity of separable biochemical species from hyperspectral microspectroscopy, and extraction of their representative spectra. The algorithm is benchmarked with established methods from satellite remote sensing, spectral unmixing, and clustering. To demonstrate the widespread applicability of the developed algorithm, we collected hyperspectral datasets using spontaneous Raman, Coherent Anti-stokes Raman Scattering and Fourier Transform IR, of seven reference compounds, an oil-in-water emulsion, and tissue-engineered extracellular matrices on poly-L-lactic acid and porcine jejunum-derived small intestine submucosa scaffolds seeded with bovine chondrocytes. We show the potential of the developed algorithm by consolidating hyperspectral molecular information with sample microstructure, pertinent to fields ranging from gastrophysics to regenerative medicine.
\end{abstract}


\section{Introduction}
The increasing amounts of data generated in chemical and biochemical experiments underpin the pivotal role of data mining techniques and algorithms. Thus, perhaps unsurprising, chemometrics has become indispensable for clustering and pattern recognition in biochemical data, especially with respect to spectroscopy and hyperspectral imaging. The field of hyperspectral imaging accelerated in the second half of the 1980's in response to the development of spaceborn imaging spectrometers \citep{Goetz1985,Goetz1987,Goetz2009}, but the associated algorithms have to date also found applicability in fields as food quality and agriculture \citep{ElMasry2007,Okamoto2009,Qiao2007}, medicine \citep{Akbari2012,Cancio2006,Zuzak2007} and art conservation \citep{Ricciardi2012,Rosi2013}. More recently, hyperspectral imaging and analysis has been adapted to chemical microspectroscopy imaging methods, such as Fourier-transform IR and Raman microspectroscopy \citep{Lau2011,Hedegaard2011,Bergner2013,Hedegaard2016,Kallepitis2017}. 

\begin{figure*}[h!]
	\centering
	\includegraphics[width=17.4cm]{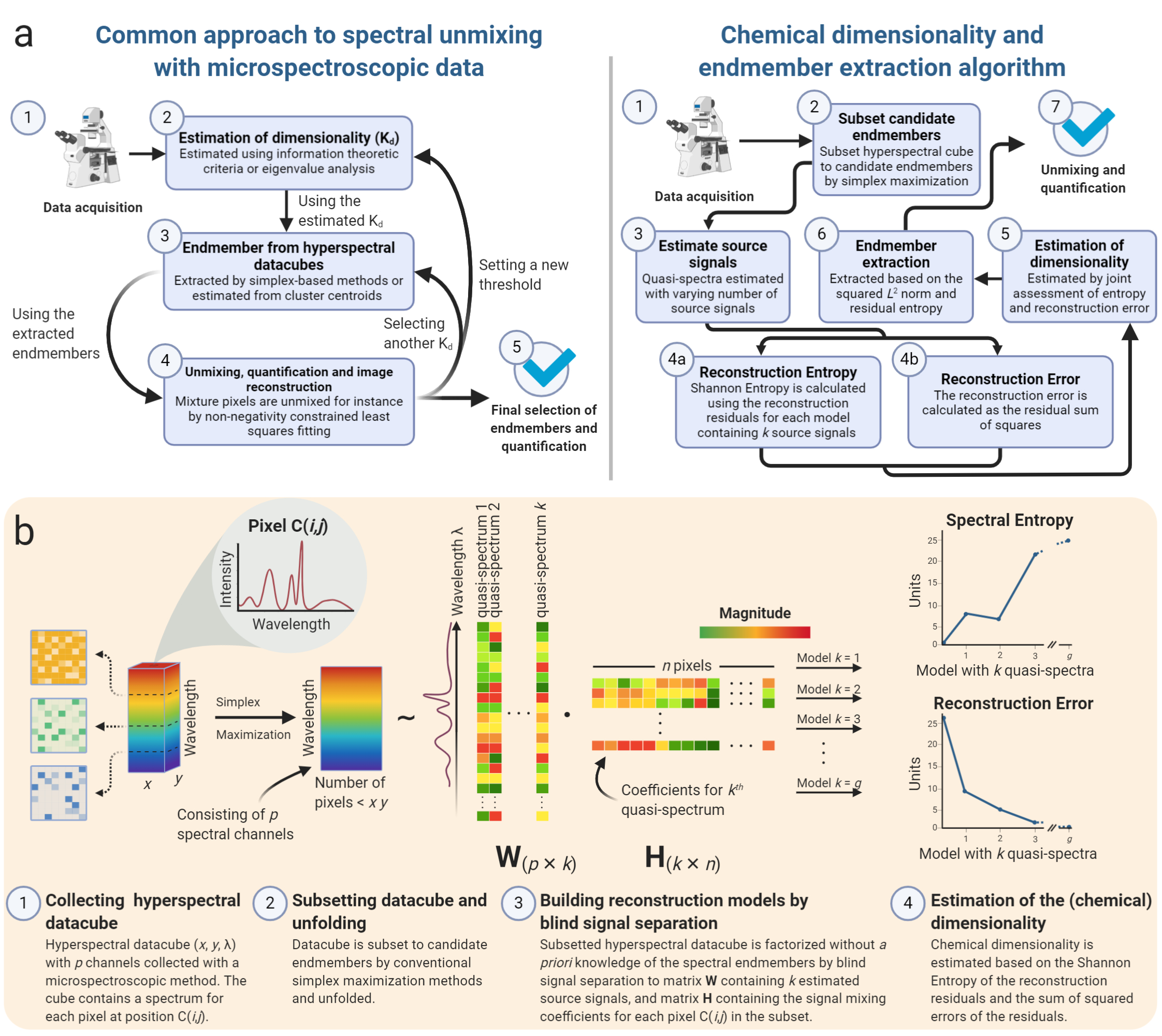}
	\caption[Comparison of hyperspectral unmixing and chemical dimensionality]{\textbf{(a)} Left: common method of data mining in hyperspectral microspectroscopy datasets. Approximate number of endmembers is determined, typically using thresholding methods. Endmember extraction is often  repeated in an exploratory fashion, making it prone to over-interpretation of returned endmembers or overlooking distinct biochemistries. Right: Algorithmic overview of the Chemical Dimensionality framework. Detailed algorithm flowchart is shown in Supplementary Figures S1 and S2, along with a full mathematical description in the methods section. Estimation of dimensionality and extraction of endmembers is carried out using information entropy and sum of squared residuals. Returned endmembers can be used to generate image reconstructions for the separated biochemistries.  \textbf{(b)} Schematic of the developed dimensionality estimation algorithm. A hyperspectral microspectroscopy image is subset to a candidate endmember matrix by repeated hyperdimensional simplex maximization. A number of blind signal separation models are built assuming an increasing number of source signals, with subsequent estimation of the Chemical Dimensionality based on  information entropy and sum of square residuals of model reconstructions.}
	\label{fig:Ifigure1}
\end{figure*}

An array of multivariate methods and chemometric algorithms have been developed for dimensionality reduction of hyperspectral datasets and extraction of representative feature vectors through data-driven approaches. These techniques include clustering samples/pixels according to feature similarity measures, such as in K-means and hierarchical clustering, or $p$-dimensional simplex algorithms, such as the N-FINDR \citep{Winter1999} algorithm. The techniques are generally applied to 1) uncover the number of distinct (bio)chemical species in a dataset, 2) extract their representative hyperspectral vectors, and 3) render chemically-specific image reconstructions. These concepts are intrinsically coupled in that extraction of their representative feature vectors, termed endmembers, typically requires setting the quantity of endmembers to extract in the first place. Even so, there is no guarantee that the first $k$ endmembers returned are indeed all endmember species. A classic example of this, applying both to clustering and simplex maximization techniques, is that even if the dataset is known to contain $k$ endmembers, it may require setting a higher $k$, to uncover the distinct endmembers. Thus, biochemical data mining in hyperspectral datasets become exploratory in nature (Figure \ref{fig:Ifigure1}), prone to over-interpretation of returned endmembers, or overlooked biochemistries. 

Herein we developed an unsupervised algorithm for (i) determining the number distinct and spectrally separable biochemistries contributing to the observed spectral variation in hyperspectral datasets, which we have termed Chemical Dimensionality (CD-DE), and (ii) extraction of endmember spectra according to the estimated dimensionality, also available as a stand-alone extraction algorithm (CD-EEA). We demonstrate widespread applicability of our method, and notable performance as a biochemical data mining and image reconstruction technique, by (i) benchmarking against pseudo-randomly generated synthetic datasets with known dimensionality, (ii) comparison of estimated Chemical Dimensionality ($k_{\text{CD}}$) to established Virtual Dimensionality estimation methods developed for remote sensing applications, and (iii) benchmarking performance influenced by varying signal-to-noise ratios and sample size. We further demonstrate versatility by (i) imaging samples ranging from liquid immiscibility in oil-in-water emulsions to extracellular matrix regeneration on biomaterial scaffolds for tissue engineering \textit{in vitro} and (ii) using a plurality of imaging techniques including Coherent Anti-stokes Raman Scattering (CARS) and Fourier Transform Infrared microspectroscopy (FTIR). Last, we compare the hyperspectral endmembers and their associated image reconstructions extracted by CD-EEA against K-means clusters and endmembers extracted by N-FINDR \citep{Winter1999}. Given that input to the Chemical Dimensionality framework can range from hyperspectral datacubes to sample feature vectors, we see the algorithm as a promising candidate for any type of multidimensional systems such as chromatography or gene expression data.

\section{Materials and Methods}
\subsection{Description of the Chemical Dimensionality Algorithm}
\subsubsection*{Chemical Dimensionality algorithm}
The algorithm is divided into two parts, a model selection algorithm that determines the Chemical Dimensionality (CD) of the input dataset, and an associated endmember extraction algorithm which uses the CD framework to extract prospect endmembers using a least square-error/maximum entropy approach. Here we first treat the model selection algorithm that determines the chemical dimensionality of a hyperspectral dataset. A flowchart is shown in Supplementary Figure S1.

The CD algorithm considers both datamatrices from spectroscopy experiments, and datacubes from microspectroscopy imaging experiments. Standard datamatrices has dimensions ($n_n$, $n_{\nu}$), where subscript $n$ denotes samples rather than pixels with an explicit spatial connotation, and $\nu$ denotes an axis of wavenumbers. As the algorithm works on datamatrices, no further preprocessing is needed prior initialization, and the dataset is directly assigned as an input datamatrix \textbf{Z}. Hyperspectral datacubes (HSDC) has dimensions ($n_x$, $n_y$, $n_\nu$), for which each pixel ($x$, $y$) in the scene contains the associated spectrum $\vec{p}(x,y)$ of length $n_\nu$. We use the term unfolding, confer earlier studies \citep{Hedegaard2016}, to describe the conversion from 3D hyperspectral datacubes to a datamatrix ready for computation, by reshaping the hyperspectral datacubes such that the new input datamatrix \textbf{Z} has dimensions ($n_x n_y$, $n_\nu$). For the purpose of image reconstruction, it is important to note the original location of each pixel from the hyperspectral datacube in the new datamatrix \textbf{Z}. In the following, we consider the input datamatrix $\textbf{Z}_{(n \times p)}$, from an unfolded HSDC if necessary, containing $n$ samples/pixels each with $p$ wavenumbers/channels. For optimum performance the input dataset $\textbf{Z}_{(n \times p)}$ should 1) contain at least one instance of a spectrally ``pure'' endmember, a prerequisite of most major endmember extraction algorithms \citep{Boardman1995,Winter1999,Nascimento2005}, 2) contain two or more endmembers, and 3) not contain endmembers that are linearly dependent, i.e. no endmembers should be expressible as a linear combination of any other set of endmembers.

\subsubsection*{Simplex maximization}
While hyperspectral image maps from spontaneous Raman and FTIR microspectroscopy are typically relatively small, in the order of hundreds to thousands of pixels in the image, other techniques, such as Coherent Anti-stokes Raman Scattering frequently yield larger images, in the order of hundreds of thousands pixels in the image. We therefore start by subsetting the input datamatrix $\textbf{Z}_{(n \times p)}$ to a reduced datamatrix $\textbf{V}_{(m \times p)}$ containing $m$ candidate endmembers where $m \ll n$, subject to $m \gg k$ where $k$ is the expected/ground true number of endmembers in the dataset. The N-FINDR \citep{Winter1999} algorithm operates on the principle that $k$ potential endmembers maximizes the volume of a simplex with vertices given from a dimensionality reduced hyperspectral dataset. However, given that the initial simplex is formed from a random set of samples/in the data, there is no guarantee that the algorithm consistently returns the same set of endmembers \citep{Chang2011}. Furthermore, even if the dataset is known to contain $k$ endmembers, it may require extracting more than $k$ endmembers to return ground true endmembers. Nonetheless, returned endmembers from a number of repeated simplex maximizations can be considered candidate endmembers, and therefore used to generate the reduced candidate endmember matrix $\textbf{V}_{(m \times p)}$. In principle most endmember extraction techniques can be used to generate the endmember candidate matrix \textbf{V}, or if the dataset is small, the input datamatrix $\textbf{Z}_{(n \times p)}$ can be used directly as a candidate endmember matrix. However, here we provide a full implementation, and adapt the simplex maximization technique described by Winter\citep{Winter1999}.

Let $i$ be the number of endmembers extracted using simplex maximization. The dimensionality of $\textbf{Z}_{(n \times p)}$ is reduced to $i – 1$ components using Singular Value Decomposition. Using the dimensionality reduced matrix, the endmember matrix, $\textbf{E}_i$ is initialized with $i$ random selection of pixels/samples.
\begin{align}
\mathbf{E}_i=\left[\begin{matrix}1&\cdots&1\\{\vec{e}}_1&\cdots&{\vec{e}}_i\\\end{matrix}\right]
\end{align}
Each pixel in the dataset is then substituted in the place of each prospect endmember ${\vec{e}}_i$, and retained if it further maximizes the simplex volume given by:
\begin{align}
Volume\left(\mathbf{E}_i\right)=\frac{1}{\left(i-1\right)!}abs\left(\left|\mathbf{E}_i\right|\right),\ \ \ \ \ \forall i\in\left\{2,\ 3,\ \ldots,\ g\right\}
\end{align}
This process is repeated for $i = 2, 3, \ldots, g$, where the resulting endmembers in $\textbf{E}_i$ found from maximization of $Volume(\textbf{E}_i)$ is continuously placed in the endmember candidate matrix $\textbf{V}_{(m \times p)}$. Owing to later comparison between the endmember candidates, it is required that $g-1 > k$.  We found that repeating the simplex maximization up to extraction of $g = 20$ endmembers generally is suitable for microspectroscopy, yielding a candidate endmember matrix $\mathbf{V}_{(m\ \times\ p)}=\mathbf{V}_{(\sum_{n=2}^{g}{n)}\ \times\ p}=\mathbf{V}_{(209\ \times\ p)}$. For large input data matrices $\textbf{Z}_{(n \times p)}$, generation of the candidate endmember matrix by repeated simplex maximization can advantageously be implemented using parallel computing. 
\subsubsection*{Self-modelling Curve Resolution using semi-nonnegativity constrained matrix factorization}
Having repeated simplex maximization to generate the candidate endmember matrix $\textbf{V}_{(m \times p)}$, the $k$ ground true endmembers are to be found among the $m$ candidates in the matrix. However, with no preconception as to the magnitude of $k$ and the corresponding endmember spectra, we can treat it as a blind signal separation problem. A canonical assumption is that endmembers
$\mathbf{E}=\left[\begin{matrix}{\vec{e}}_1&{\vec{e}}_2&\ldots&{\vec{e}}_k\\\end{matrix}\right]$, are considered vectors in a real vector space and describe mixture voxels, pixels or samples through convex mixing. This implies that, for each voxel, pixel or sample with spectrum ${\vec{p}}_{(1\times p)}$ in the unfolded data input matrix:
\begin{multline}
\label{eq:eq42}
{\vec{p}}_{i_{\left(1\times p\right)}}=\left(\mathbf{E}_{\left(p\times k\right)}\ {\mathbf{x}_i}_{\left(k\times1\right)}\right)^T+{\vec{\epsilon}}_i,\ \ \ \ \ \forall i\in{1,\ 2,\ \ldots,n}\ \\ s.t.\ \ \mathbf{x}_i\geq0
\end{multline}
Where $x$ can be thought of as a vector of mixing weights for the convex mixing:
\begin{align}
\mathbf{x}_i=\left[\begin{matrix}w_{i1}&w_{i2}&\cdots&w_{ik}\\\end{matrix}\right]^T,\ \ \ \ \ s.t.\ \ \sum_{j=1}^{k}w_{ij}=1
\end{align}
While the sum-to-one constraint on linear mixing models is reasonable, it is an unnecessary constraint when determining the Chemical Dimensionality. Furthermore, not all microspectroscopy techniques lend themselves equally well to quantitative, idealized mixing models, for instance due to an inherent non-resonant background such as in conventional Coherent Anti-stokes Raman Scattering (CARS) imaging. If the sum-to-one constraint is dropped, estimating ${\vec{p}}_{i_{\left(1\times p\right)}}$ becomes a conical mixing problem, retaining the non-negativity constraint on $\textbf{x}_i \geq 0$. The mixing weights in $\textbf{x}$ can then be obtained by solving a non-negative constrained minimization problem in a least-squares sense:
\begin{align}
\label{eq:eq45}
\min_x \left\Vert \textbf{E} \textbf{x}_i - p^{\text{T}}_i \right\Vert^{2}_{\text{2}},\ \ \ \ \ \forall i\in\left\{1, 2,\ \ldots,\ n\right\} \ \ \ \ s.t. \ \  \textbf{x}_i \geq 0 
\end{align}
Where $\Vert \cdot \Vert_2$ denotes the $\ell^2$ norm. To determine the Chemical Dimensionality we let \textbf{E} be of size $(p \times u)$, i.e. $u$ feature vectors with data collected at $p$ wavenumbers/channels. As the signature of the $u$ feature vectors is unknown, determining \textbf{E} becomes a blind signal separation problem. Here, blind signal separation was carried out using non-negative matrix factorization (NMF). At its core, non-negative matrix factorization seeks to express a non-negative matrix $\mathbf{X}\in\mathbf{R}^{p\times n}$ in terms of $\mathbf{W}\in\mathbf{R}^{p\times k}$ and $\mathbf{H}\in\mathbf{R}^{k\times n}$, such that:
\begin{align}
\label{eq:eq46}
\mathbf{X}\approx\mathbf{WH},\ \ \ \ \ s.t.\ \ \mathbf{W},\ \mathbf{H}\geq0
\end{align}
In Eq. \ref{eq:eq46}, \textbf{X} can be thought of as the input spectral data matrix, \textbf{W} a set of $u$ feature vectors i.e. ``quasi-endmembers'', and \textbf{H} containing the weights of each quasi-endmember. Depending on the type of spectroscopy used for data acquisition, candidate endmembers may contain slightly negative arbitrary intensities, for instance owing to prior pre-processing with baseline correction algorithms, challenging the non-negativity constraint on \textbf{X}. We therefore adopted the more lenient semi-NMF method by Ding et al.\citep{Ding2010}, in which \textbf{W} is unconstrained and \textbf{H} is strictly non-negative. The semi-NMF optimization problem is thus described by:
\begin{align}
\label{eq:eq47}
\min_\text{\textbf{W,H}} \frac{1}{2} \left\Vert \text{\textbf{X}}-\text{\textbf{WH}}\right\Vert^{2}_{\text{F}}, \ \ \ \ \ s.t \ \ \text{\textbf{H}} \geq 0
\end{align}
Where $\Vert \cdot \Vert_\text{F}$ denotes the Frobenius norm. In this Chemical Dimensionality algorithm we used the semi-NMF method as implemented by Li and Ngom\citep{Li2013}, using the fast combinatorial non-negative least squares algorithm developed by van Benthem and Keenan\citep{VanBenthem2004}. From Eq. \ref{eq:eq47} we generate $g$ models consisting of $u$ feature vectors where $u = 1, 2, \ldots, g$, as for the repeated simplex maximization, and \textbf{X} is the candidate endmember matrix $\mathbf{V}_{(\sum_{n=2}^{g}{n)}\ \times\ p}$, such that:
\begin{multline}
\min_\text{\textbf{W,H}} \frac{1}{2} \left\Vert \textbf{V}^{\text{T}} - \textbf{W}_{(p \times u)} \textbf{H}_{(u \ \times \ \sum_{n=2}^{g}{n})}  \right\Vert^{2}_{\text{F}},\ \\ \forall u\in\left\{1, 2,\ \ldots,\ g\right\}  \ \ s.t. \ \ \textbf{H} \geq 0
\end{multline}

\subsubsection*{Error reduction}
To select the final model of $\mathbf{W}\in\mathbf{R}^{p\times u}$ and $\mathbf{H}\in\mathbf{R}^{u\times n}$, where $u$ becomes the Chemical Dimensionality, we adapted the Shannon Entropy \citep{Shannon1948}, a concept from information theory, to calculate a total entropy of the reconstruction residuals. We also calculated an error reduction  term for each of the $u$ blind signal separation models, and together, the two metrics were used to determine the Chemical Dimensionality. Firstly, for both metrics, let $\textbf{R}_u$ contain the reconstruction residuals from the semi-nonnegative matrix factorization with $u$ feature vectors:
\begin{align}
\mathbf{R}_u=\ \mathbf{V}_{\left(m\ \times\ p\right)}-\left(\mathbf{W}_{\left(p\times u\right)}\mathbf{H}_{\left(u \times m \right)}\right)^\text{T}\ \ \ \ \ \ \forall u\in{1,\ 2,\ \ldots,g}
\end{align}
We defined the error reduction for model $u$ as the ratio between the reduction in normalized sum of square error for models $u-1$ to $u$, and $u$ to $u+1$ $\forall u=2,\ 3\ ,\ \ldots,g-1$. Alternatively, the error reduction in this sense can be thought of as measure of ``sudden'' improvements in how well a semi-NMF model with $u$ feature vectors describes the candidate endmember matrix $\mathbf{V}_{(m\ \times\ p)}$.  Denoting the matrix elements of $\textbf{R}_u$ as $\textbf{R}_{u,mp}$, the sum of square error for the $u^{th}$ semi-NMF model, $s_u$, is given in Eq. \ref{eq:eq410}, which for interpretability is normalized, such that $\varepsilon_u\in\left[0,1\right]$. Noting that $s_u\in\mathbf{R}^+\cup\left\{0\right\}\ \ \forall u\in{1,\ 2,\ \ldots,g}$ should be monotonically decreasing as $u$ goes from $1$ to $g$:
\begin{align}
\label{eq:eq410}
s_u=\sum_{m,p}\mathbf{R}_{u,mp}^2\ \ \ \ \ \ \forall u\in{1,\ 2,\ \ldots,g}
\end{align}
\begin{align}
\epsilon_u=\frac{1}{\max_{1\le i\le g}{s_i}}\ s_{u\ \ \ \ \ }\forall u\in\left\{1,\ 2,\ \ldots,g\right\}
\end{align}
As $u$ varies from $1$ to $g$ in steps of $1$, we calculate the  error reduction, $\rho\in\mathbf{R}^+\cup\left\{0\right\}$, for model $u$ by:
\begin{align}
\rho\left(u\right)=\frac{\epsilon_{(u-1)}-{\ \epsilon}_u\ }{\epsilon_u-\epsilon_{(u+1)}}\ \ \ \ \ \ \forall u\in\left\{2,\ 3,\ \ldots,g-1\right\}
\end{align}

\subsubsection*{Entropy of reconstruction}
Subsequently, we calculated the information theoretic metric based on the Shannon entropy \citep{Shannon1948}, which hereinafter will be referred to as the reconstruction entropy. The reconstruction entropy is given by: 
\begin{align}
\label{eq:eq413}
\text{H}(b_i)=\ -\sum_{i}{{b}_i\ln{b_i}}
\end{align}
Inspired by prior use of Shannon entropy \citep{Widjaja2003}, we calculated $b_i$ such that $b_i\ \forall i$ sums to unity. Let ${\vec{\mathbf{r}}}_{c,u}=\left[\begin{matrix}r_{c,u,1}&r_{c,u,2}&\cdots&r_{c,u,p}\\\end{matrix}\right]^\text{T}$, where $c\in\left\{1,\ 2,\ \ldots,m\right\}$, be a residual vector of $\textbf{R}_u$ with dimensions $(p \times 1)$ across the associated wavenumbers $\vec{\mathbf{v}}=\left[\begin{matrix}\nu_1&\nu_2&\cdots&\nu_p\end{matrix}\right]^\text{T}$, then:
\begin{multline}
\label{eq:eq414}
{b({\vec{\mathbf{r}}}_{c,u},\vec{\mathbf{v}})}_i=\ \frac{\left|\frac{r_{c,u,i}\ -{\ r}_{c,u,(i-1)}\ }{\nu_i-\ \nu_{(i-1)}}\right|}{\sum_{h\in \left\{2,\ \ 3,\ \ \ldots,\ p\right\}}\left|\frac{r_{c,u,h}\ -{\ r}_{c,u,(h-1)}}{\nu_h - \nu_{(h-1)}}\right|}\ \ \ \\ \forall i\in\left\{2,\ \ 3,\ldots,\ p\right\} 
\end{multline}

Where for consistency ${b({\vec{\mathbf{r}}}_{c,u}\ ,\vec{\mathbf{v}})}_i=0\Rightarrow\ {b({\vec{\mathbf{r}}}_{c,u}\ ,\vec{\mathbf{v}})}_i\ln{{b({\vec{\mathbf{r}}}_{c,u}\ ,\vec{\mathbf{v}})}_i}=\ 0$. When the elements of ${b({\vec{\mathbf{r}}}_{c,u}, \vec{\mathbf{v}})}_i$ are numerically equivalent, H is maximized. Correspondingly, the reconstruction entropy will tend to increase as the residuals tend to identically distributed noise, and decrease if the residuals carry spectral information. For a given semi-NMF model with $u$ feature vectors, we calculate a total entropy score $S_u\in\mathbf{R}^+$ by summing the entropy from Eq. \ref{eq:eq413} for all $c \in {1,\ 2,\ \ldots,m}$ residual vectors in $\textbf{R}_u$, which is then repeated for $\forall u\in\left\{1,\ 2,\ \ldots,g\right\}$. 
\begin{align}
S_u=\ \sum_{c\in\left\{1,\ 2,\ \ldots,m\right\}}{\text{H}\left(b({\vec{\mathbf{r}}}_{c,u}\ ,\vec{\mathbf{v}})\right)\ \ \ \ \ \ \forall u\in\left\{1,\ 2,\ \ldots,g\right\}}
\end{align}

\subsubsection*{Estimation of the Chemical Dimensionality}
The reconstruction entropy is expected to increase, as the reconstruction residuals tend to identically distributed random noise. Consequently, maximizing the reconstruction entropy may not (necessarily) estimate a meaningful dimensionality. Instead, the chemical dimensionality, $k_{\text{CD}}$, is estimated by searching for an increase in reconstruction entropy at or in the vicinity of the $u^{th}$ semi-NMF model, providing the maximum error reduction.  
\begin{align}
z=\ \arg{\max_{u\in\left\{2,\ 3,\ \ldots,g-1\right\}}{\rho\left(u\right)}}
\end{align}
If model $z$ was also associated with an increase in reconstruction entropy, $S_z-S_{\left(z-1\right)}>0$ then $ k_{\text{CD}}= z$. Otherwise, we seek the nearest model $z\pm i,\ \ i\in\left\{1,\ 2,\ 3,\ ...\right\}$, subject to $ (z\pm i)\in\mathbf{N}\ |\ 2\le z\pm i\le g-1$, associated with an increase in reconstruction entropy $S_{(z\pm i)}-S_{\left(z\pm i-1\right)}>0$. If $S_{(z+i)}-S_{\left(z+i-1\right)}>0$ (while ${S}_{(z-i)}-S_{\left(z-i-1\right)}\le 0$) is the nearest model associated with an increase in reconstruction entropy, then $k_{\text{CD}}= z+i$, and conversely $k_{\text{CD}}= z-i$ if $S_{(z-i)}-S_{\left(z-i-1\right)}>0$ (while $S_{(z+i)}-S_{\left(z+i-1\right)}\le 0$) is nearest $z$. In case both $S_{(z-i)}-S_{\left(z-i-1\right)}>0$ and $S_{(z+i)}-S_{\left(z+i-1\right)}>0$ for a given $i$, then we seek the model that resulted in the greatest increase in reconstruction entropy, i.e. $k_{\text{CD}}= \arg{\max_{f\in{z-i,z+1}}{S_f-S_{\left(f-1\right)}}}$.
\subsubsection*{Chemical Dimensionality endmember extraction}
After estimation of the Chemical Dimensionality, $k_{\text{CD}}$, the algorithm proceeds to extract $k_{\text{CD}}$ samples/pixels from the endmember candidate matrix identified as actual  endmembers of the dataset, according to the $\ell^2$ norm of the reconstruction residuals and the reconstruction entropy. First, we define the Chemical Dimensionality endmember matrix $\textbf{E}_{\text{CD}}$, containing $k_{\text{CD}}$ endmembers, such that:
\begin{multline}
\label{eq:eq417}
\mathbf{E}_{CD,\ \ {\ \ \ k}_{CD}\times p}=\left[\begin{matrix}{\vec{e}}_1&{\vec{e}}_2&\cdots&{\vec{e}}_{k_{\text{CD}}}\\\end{matrix}\right]^\text{T} \\ s.t.{\ \vec{e}}_1\neq{\vec{e}}_2\neq\ldots\neq{\vec{e}}_{k_{\text{CD}}}
\end{multline}
It should be noted that, as a consequence of the repeated simplex maximization in Eq. \ref{eq:eq42}, the endmember candidate matrix may contain multiple occurrences of the same endmember candidate. For the initialization in Eq. \ref{eq:eq417}, $\textbf{E}_{\text{CD}}$ is consequently populated with the $k_{\text{CD}}$ samples/pixels in the endmember candidate matrix $\mathbf{V}_{m\ \times\ p}$ with the lowest $\ell^2$ norm, subject to ${\vec{e}}_1\neq{\vec{e}}_2\neq\ldots\neq{\vec{e}}_{k_{\text{CD}}}$. This condition must also be checked when the algorithm proceeds to iteratively swap endmembers in $\textbf{E}_{\text{CD}}$ with new prospect endmembers. 

Having populated the initial endmember matrix $\textbf{E}_{\text{CD}}$, the non-negativity constrained least squares minimization problem in Eq. \ref{eq:eq45} is solved using $\mathbf{E}=\mathbf{E}_{CD}^\text{T}$ and ${\ \vec{p}}_i$ as the $i^{th}$ candidate endmember in $\mathbf{V}_{(m\ \times\ p)},\ \ \forall i\in{1,\ 2,\ \ldots,\ m}$.
\begin{align}
\label{eq:eq418}
\min_x \left\Vert \textbf{E}^{\text{T}}_{\text{CD}} \textbf{x}_i - p^{\text{T}}_i \right\Vert^{2}_{\text{2}},\ \ \ \ \ \forall i\in\left\{1, 2,\ \ldots,\ n\right\} \ \ s.t. \ \ \textbf{x}_i \geq 0 
\end{align}

Having solved for $\textbf{x}$, the performance of endmembers initially present in $\textbf{E}_{\text{CD}}$ in terms of the $\ell^2$ norm is then given by:
\begin{align}
P_{L^2} = \sum_{i\in\left\{1, 2,\ \ldots,\ m\right\}} \left\Vert \textbf{E}^{\text{T}}_{\text{CD}} \textbf{x}_i - p^{\text{T}}_i \right\Vert^{2}_{\text{2}}
\end{align}

To calculate the reconstruction entropy of the endmembers, let: 
\begin{align}
{\vec{\mathbf{r}}}_c={\mathbf{E}_{CD}^T\ \mathbf{x}_c-{\vec{p}}_c^T\ =\left[\begin{matrix}r_{c,1}&r_{c,2}&\cdots&r_{c,p}\\\end{matrix}\right]}^\text{T}
\end{align}

\noindent be a residual vector from the minimization problem in Eq. \ref{eq:eq418}, where $c\in\left\{1,\ 2,\ \ldots,m\right\}$ and ${\vec{\mathbf{r}}}_c$ has dimensions $(p \times 1)$ across the associated wavenumbers $\vec{\mathbf{v}}=\left[\begin{matrix}\nu_1&\nu_2&\cdots&\nu_p\\\end{matrix}\right]^\text{T}$, then ${b({\vec{\mathbf{r}}}_c\ ,\vec{\mathbf{v}})}_i$ is calculated using Eq. \ref{eq:eq414}, and the reconstruction entropy becomes:
\begin{align}
\label{eq:eq420}
P_S=\ \sum_{c\in\left\{1,\ 2,\ \ldots,m\right\}} \text{H}\left(b({\vec{\mathbf{r}}}_c\ ,\vec{\mathbf{v}})\right)
\end{align}
The algorithm loops through each of the m candidate endmembers in \textbf{V}, which is then substituted in place of each ${\vec{e}}_j,\ j\in{1,\ 2,\ \ldots,\ k_{\text{CD}}}$. For each swap, Eq. \ref{eq:eq418}-\ref{eq:eq420} is calculated. If swapping one or more endmembers in $\mathbf{E}_{\text{CD}}$ with a given candidate endmember from \textbf{V} results in a reconstruction $\ell^2$ norm lower than $P_{\ell^2}$ , then the candidate endmember is swapped in place of the endmember in $\mathbf{E}_{\text{CD}}$ for which the reconstruction entropy, $P_S$, was maximized compared to the other swaps, regardless of whether the swap also resulted in the minimum reconstruction $\ell^2$ norm or not. Subsequently, $P_{\ell^2}$  is updated with the reconstruction $\ell^2$ norm of the new endmember matrix $\mathbf{E}_{\text{CD}}$. The procedure is continued until looping through the candidate endmember matrix \textbf{V} no longer results in any swaps. A flowchart of the endmember extraction algorithm can be found in Supplementary Figure S2.

\subsection{Methodological procedures}
\subsubsection*{Generated datasets}
To benchmark the Chemical Dimensionality algorithm we generated a series of datasets with 1) a known number of endmembers ranging from 2 to 10 endmembers, each with 2) random convex mixtures of the 9 endmember sets, to a total of 300, 1000 and 5000 samples, and 3) random convex mixtures of endmember sets to a total of 5000 samples with signal-to-noise ratios (SNR) ranging from 1000:1 to 10:1. New endmembers were generated for each endmember sets $k = 2, \ 3, \ \ldots, \ 10$. For an endmember set with $k$ endmembers, each of the $k$ pseudo-spectra were generated by mixing a random number $f = 1,\ 2,\ \ldots, \ 6,$ of Gaussian functions each with random height ($a$), width ($s$) and peak center on the spectral axis ($m$) (Eq. \ref{eq:eq421}). To test the algorithm specifically on its performance on datasets of various sample sizes and signal-to-noise ratios, all endmembers were vector normalized to unit length prior mixing, to avoid confounding with scale. Although arbitrary, we let the x-axis range from $\nu_1= 900$ to $\nu_{1001}= 1900$, i.e. consisting of 1001 channels/wavenumbers. 
\begin{multline}
\label{eq:eq421}
e_i(x)=\ \sum_{j=1}^{f}{a_jexp{\left(-\frac{\left(x-m_j\right)^2}{2s_j^2}\right),}}\ \ \ \ \forall x\in\left\{\nu_1,\ \nu_2,\ \ldots\ \nu_h\right\},\\ \forall i\in{1,\ 2,\ \ldots,k}
\end{multline}
\noindent Where:
	\begin{align}
	a_j\ & \sim U\left(\left[0,1\right]\right) \\ s_j &\sim  U\left(\left[\nu_1,\nu_h\right]\right)\ \forall\ s_j\in\mathbf{N} \\ m_j &\sim  U\left(\left[0,50\right]\right)
	\end{align}

Once the 9 endmember sets were prepared, the endmembers of each set were mixed such that $\sum_{j=1}^{k}w_j=1$, where weights $w_j$ were pseudo-randomly generated to mix a total of $300 - k$, $1000 - k$ and $5000 - k$ spectra. The $k$ original endmembers were then appended to the data matrix, and the entire data matrix was shuffled. The final locations of the $k$ original endmembers in the data matrix were therefore unknown.

Prior running the Chemical Dimensionality algorithm on the synthesized datasets, white Gaussian noise was added to each sample in each mixture dataset, to avoid strict linear dependency. For studying the effect of sample size on the CD algorithm, all mixture datasets for each endmember set, was added noise to an SNR of 1000:1, while the mixture datasets containing 5000 samples were also added white Gaussian noise to an SNR of 100:1, 50:1, 20:1 and 10:1, to test the performance of the algorithm with increasing noise levels (Table \ref{tab:ItableS1}).
\begin{table}[t!]
	\small
	\renewcommand{\arraystretch}{1.0}%
	\centering
	\caption[Synthesized datasets with varying endmembers, sample mixtures and SNRs]{Overview of synthesized datasets with varying endmember constituents, sample mixtures and signal-to-noise ratios.}
	\label{tab:ItableS1}
	\begin{tabular}[c]{p{2cm}p{2cm}l}
		\toprule
		\textbf{Number of endmembers}                                                           & \textbf{Samples in mixture dataset}                                                                                                                                                                                                                   & \textbf{Signal-to-noise ratio}      \\ \midrule                                                                                                                                                                                                                                                                                
		2 & \begin{tabular}[c]{@{}l@{}}300\\ 1000\\ 5000\end{tabular} & \begin{tabular}[c]{@{}l@{}}1000:1\\ 1000:1\\ 1000:1, 100:1, 50:1, 20:1, 10:1\end{tabular}  \\ 
		& &  \\
		3 & \begin{tabular}[c]{@{}l@{}}300\\ 1000\\ 5000\end{tabular} & \begin{tabular}[c]{@{}l@{}}1000:1\\ 1000:1\\ 1000:1, 100:1, 50:1, 20:1, 10:1\end{tabular}  \\
		& &  \\
		4 & \begin{tabular}[c]{@{}l@{}}300\\ 1000\\ 5000\end{tabular} & \begin{tabular}[c]{@{}l@{}}1000:1\\ 1000:1\\ 1000:1, 100:1, 50:1, 20:1, 10:1\end{tabular}  \\
		& &  \\
		5 & \begin{tabular}[c]{@{}l@{}}300\\ 1000\\ 5000\end{tabular} & \begin{tabular}[c]{@{}l@{}}1000:1\\ 1000:1\\ 1000:1, 100:1, 50:1, 20:1, 10:1\end{tabular}  \\
		& &  \\
		6 & \begin{tabular}[c]{@{}l@{}}300\\ 1000\\ 5000\end{tabular} & \begin{tabular}[c]{@{}l@{}}1000:1\\ 1000:1\\ 1000:1, 100:1, 50:1, 20:1, 10:1\end{tabular}  \\
		& &  \\
		7 & \begin{tabular}[c]{@{}l@{}}300\\ 1000\\ 5000\end{tabular} & \begin{tabular}[c]{@{}l@{}}1000:1\\ 1000:1\\ 1000:1, 100:1, 50:1, 20:1, 10:1\end{tabular}  \\
		& &  \\
		8 & \begin{tabular}[c]{@{}l@{}}300\\ 1000\\ 5000\end{tabular} & \begin{tabular}[c]{@{}l@{}}1000:1\\ 1000:1\\ 1000:1, 100:1, 50:1, 20:1, 10:1\end{tabular}  \\
		& &  \\
		9 & \begin{tabular}[c]{@{}l@{}}300\\ 1000\\ 5000\end{tabular} & \begin{tabular}[c]{@{}l@{}}1000:1\\ 1000:1\\ 1000:1, 100:1, 50:1, 20:1, 10:1\end{tabular}  \\
		& &  \\
		10 & \begin{tabular}[c]{@{}l@{}}300\\ 1000\\ 5000\end{tabular} & \begin{tabular}[c]{@{}l@{}}1000:1\\ 1000:1\\ 1000:1, 100:1, 50:1, 20:1, 10:1\end{tabular}  \\\bottomrule                                                                                                                                           
	\end{tabular}
\end{table}

In this work we calculated signal/spectral strength as the root-mean-square of the signal. Consider a pixel/sample $p_i$ with the associated spectrum ${\vec{p}}_i\left(\nu\right)_{(1\times p)}=\left[\begin{matrix}\nu_1&\nu_2&\ldots&\nu_p\\\end{matrix}\right]$. Signal strength $\text{E}_{\text{RMS}}$ is then given by:
\begin{align}
\text{E}_{\text{RMS}}=\sqrt{\frac{\sum_{j=1}^{p}\nu_j}{p}}
\end{align}
The signal strength $\text{E}_{\text{RMS}}$ was calculated for each of the samples in each of the mixture datasets. The weakest signal was used to determine the amount of white Gaussian noise to be added to obtain the desired signal-to-noise ratio. The same amount of noise was added to all remaining samples in the dataset, such that the noise level in each sample at least matches the stated SNR.
\subsubsection*{Benchmarking Chemical Dimensionality to established Virtual Dimensionality methods}
The estimated Chemical Dimensionality, $k_{\text{CD}}$, was compared to the dimensionality estimated by five different methods from analytical chemistry, information theory, and Neyman-Pearson detection theory, all of which has previously been applied to hyperspectral imagery from satellite remote sensing \citep{Chang2003,Chang2004} in estimation of the Virtual Dimensionality. Benchmarking was based on well-defined synthesized datasets and was carried out for cases where 1) $p < n$, $p \approx n$, and $p > n$, where $p$ is the number of channels, and $n$ is the number of pixels/samples in the dataset, and 2) signal-to-noise ratios as follows; 1000:1, 100:1, 50:1, 20:1 and 10:1.

On the basis of Neyman-Pearson detection theory, the dimensionality of the generated datasets was estimated by the Harsanyi-Farrand-Chang method(Chein-I Chang, 2013) (HFC) using a false-alarm probability of $10^{-5}$, and from information theory the dimensionality was estimated by the Akaike Criterion (AIC) and the Minimum Description Length (MDL), as detailed in prior works \citep{Wax2009,Chang2004}. These methods ideally require that $n \gg p$, and especially for the AIC and MDL criteria must noise be independent and identically distributed \citep{Chang2004}. The Chemical Dimensionality was also compared to the dimensionality predicted by Malinowski’s Factor Indicator Function (FIF), which was developed and specifically tested on data from spectroscopy, magnetic resonance and chromatography \citep{Malinowski1977}. Last, we implemented an explained variance thresholding method which frequently appears in literature. Here, the dimensionality was estimated by finding the number of principle components for which the explained variance would describe at least \SI{95}{\percent} and \SI{99}{\percent} of the total variability, here denoted $\lambda_{95}$ and $\lambda_{99}$ respectively, by Principal Component Analysis (PCA) using Singular Value Decomposition (SVD). As these methods cannot extract the estimated number of endmembers from a dataset, the Chemical Dimensionality algorithm was benchmarked separately on its ability to extract endmembers from a hyperspectral dataset.
\subsubsection*{Benchmarking Chemical Dimensionality endmember extraction}
Endmembers extracted by the Chemical Dimensionality algorithm were benchmarked to those extracted by N-FINDR \citep{Winter1999}, a simplex maximization-based algorithm that has previously been extensively deployed for unmixing of hyperspectral data from various microspectroscopy techniques \citep{Hedegaard2011,Lau2011,Bergner2013,Hedegaard2016}. Extraction was conducted for both the synthesized datasets as well as experimental datasets, and the number of endmembers extracted was dictated by the estimated Chemical Dimensionality, $k_{\text{CD}}$. For the synthesized datasets and the spontaneous Raman spectroscopy dataset of seven compound standards, the endmembers extracted by N-FINDR and the CD-EEA were compared to the ground true, known, endmembers. For the synthesized dataset, performance was evaluated with 2, 5 and 10 known endmembers mixed to 5000 samples with SNR 1000:1. The experimental hyperspectral imaging datasets enabled image reconstruction by solving the non-negativity constrained least squares (NNLS) problem in Eq. \ref{eq:eq45}. Here, image reconstruction was carried out by NNLS of endmembers extracted by the N-FINDR and CD-EEA again using the estimated $k_{\text{CD}}$, and these were further compared with their respective cluster images obtained from K-means clustering using the K-Means++ algorithm \citep{Arthur2007}.
\subsubsection*{Computational setup}
The algorithm was implemented in a x64 version of MATLAB R2018b Update 3 (MathWorks Inc., Natick, MA, USA) on a desktop computer running Windows 10. The system consisted of 10 cores Intel® Core™ i9-9820X CPU @ 3.30GHz and 128 GB of RAM. The algorithm was implemented exploiting the benefits of parallel computing where appropriate. We ran the algorithm using $j_{max} = 5$ for $g = 2,\ 3,\ \ldots,\ 20$ simplex maximization repeats (see Supplementary Figure \ref{fig:IschematicS1}), yielding a candidate endmember matrix of 209 candidate endmembers. All 10 cores were deployed for parallel computing.
\subsection{Experimental procedures}
\subsubsection*{Spontaneous Raman spectroscopy of compound standards}
Seven compound standards were measured using spontaneous Raman spectroscopy. The standards were from a Bruker Standard Reference kit, and included; Polystyrene, Tylenol, Silicon, Calcium carbonate and Naphthalene. Stearic acid was obtained from Sigma Aldrich and \textbeta-Tricalcium phosphate was a kind donation from Particle3D. Raman spectra were collected using a Fiber Optic Probe-based setup as previously described \citep{Walther2019}. Briefly, spontaneous Raman spectra were collected using a near-infrared laser (B\&W TEK Inc., Newark, DE, USA) with excitation at \SI{785}{\nano\metre}, and a power output at \SI{50}{\milli\watt}. The Filtered Fiber Optic Probe (EmVision LLC., Loxahatchee, FL, USA) contains seven low OH collection fibers with $\text{NA} = 0.22$ and \SI{300}{\micro\metre} core, arranged in a donut-shape around the \SI{300}{\micro\metre} core ($\text{NA} = 0.22$) excitation fiber. Filtered Raman signals in the collection fibers were led to an Ibsen Eagle Raman-S OEM spectrometer (Ibsen Photonics A/S, Farum, Denmark). The Ibsen Eagle Raman-S spectrometer is based on a transmission grating design  with light projected onto an Andor iVac 316 (Andor, Belfast, Northern Ireland) NIR-optimized back-illuminated Low Dark Current Deep-depletion CCD (2000 $\times$ 256 pixels) cooled to \SI{-70}{\degreeCelsius} using the in-built thermoelectric Peltier cooler. Spectra were collected in full-vertical binning mode with 1s integration and 5 co-additions. The collected Raman spectra for each of the seven standards was subsequently baseline-corrected using a combined asymmetric least squares Whittaker smoothing method \citep{Eilers2003,Eilers2005}, with input parameters $10^{-6} \leq p \leq 10^{-3}$ and $10^2 \leq \lambda \leq 10^4$. Here explicitly $p$ is an asymmetry parameter, and $\lambda$ is a smoothing parameter \citep{Eilers2005}. 

The compound standards were used to test the performance of the algorithm with experimentally obtained spectra. Therefore, contrary to the synthesized endmember sets, the spontaneous Raman standards were not vector normalized prior mixing. The seven standards were mixed as-is, such that such that $\sum_{j=1}^{k}w_j=1$ to a set of 5000 samples, and added noise to an SNR of 1000:1 as previously described to avoid strict linear dependency.
\subsubsection*{FTIR Imaging of Polylactic acid (PLLA) and Small Intestine Submucosa (SIS) tissue scaffolds}
PLLA scaffold was prepared by dissolving PLLA in 1,1,1,3,3,3-hexafluoroisopropanol (HFIP) at \SI{8}{\percent} w/v overnight. Dissolved PLLA was subsequently electrospun into a randomly oriented scaffold consisting of \SI{1}{\micro\metre} PLLA fibers, to a mean scaffold thickness of \SI{120}{\micro\metre}. PLLA and HFIP was purchased from Sigma Aldrich. Small Intestine Submucosa (SIS) biomaterial scaffolds were derived from porcine jejunum by mechanical delamination and subsequently decellularized for two hours in \SI{0.1}{\percent} (vol/vol) peracetic acid, \SI{4}{\percent} (vol/vol) ethanol, and \SI{96}{\percent} (vol/vol) DI water. Prior final preparation of the multi-layered scaffolds, electrospun PLLA was sterilized in \SI{70}{\percent} ethanol and extensively rinsed in PBS. SIS Scaffolds were freeze-dried and sterilized with ethylene oxide, and rehydrated in sterile PBS prior layering. Five layers of each scaffold were used for layering, and these were each seeded with isolated bovine chondrocytes with 200K cells per layer. After seeding and cell adhesion, the media was replaced by a chondrogenic differentiatiation medium containing DMEM (\SI{4.5}{\gram\per\litre}~l-glucose), \SI{50}{\milli\gram\per\milli\litre} L-Proline, \SI{50}{\milli\gram\per\milli\litre} ascorbic acid, 0.1 mM sodium pyruvate, \SI{0.1}{\percent} v/v ITS premix and \SI{10}{\nano\gram\per\milli\litre} TGF-\textbeta3. All materials were procured from Sigma Aldrich, except ITS premix and TGF-\textbeta3 which were procured from BD Biosciences (HQ, NJ, USA) and Lonza (HQ, Basel, Switzerland) respectively. The multi-layered scaffolds were cultured for four weeks prior sectioning and image acquisition. Detailed descriptions on scaffold preparation and isolation of bovine chondrocytes can be found in earlier works \citep{Freytes2004,McCullen2012,Accardi2013,Davis2014,Steele2014}.

Hyperspectral maps of the incubated PLLA and SIS scaffolds were collected using a Perkin Elmer Spectrum 100 FTI-IR connected to a Perkin Elmer AutoImage microscope. Sections of the scaffolds were cut in a cryostat microtome by embedding in O.C.T., and subsequently placed on \ce{MgF2} slides. Hyperspectral data was collected with 10 integrations across the spectral region from \SIrange{1000}{4000}{\per\centi\metre}, at a spatial resolution of \SI{20}{\micro\metre}. Spectral data in the region from \SIrange{1000}{1860}{\per\centi\metre} was used for further analysis.
\subsubsection*{Coherent Anti-stokes Raman Scattering (CARS) microspectroscopy of immiscible lipid-water phases in mayonnaise}
Immiscibility study of the lipid and aqueous phases in mayonnaise using the Chemical Dimensionality and endmember extraction algorithm was conducted using Coherent Anti-stokes Raman Microspectroscopy. The imaging setup included a Leica TCS SP8 inverted microscope (Leica Microsystems GmbH, Wetzlar, Germany) equipped with a picoEmerald pico-second pulsed laser (APE GmbH,  Berlin, Germany) delivering a \SI{300}{\milli\watt} Stokes beam at \SI{1064}{\nano\metre}, and a tunable optical parametric oscillator (OPO) delivering a pump/probe beam in the range \SIrange{780}{940}{\nano\metre} with \SI{300}{\milli\watt}. The Stokes and pump/probe lasers were focused onto the sample using an HC PL IRAPO 40x/1.1 objective, and a spectral stack was collected using the non-descanned detectors in forward mode by tuning the pump/probe laser in the range from \SIrange{794.6}{826.6}{\nano\metre} in steps of \SI{1}{\nano\metre} ($\sim$\SIrange{3200}{2713}{\per\centi\metre}). Handmade mayonnaise was prepared at the SDU FoodLab, and kindly donated by Mathias Porsmose Clausen.

\section{Results}
\label{sec:Iresults}

\subsection{Overview of the CD framework}

A detailed flowsheet of the Chemical Dimensionality framework, CD-DE and CD-EEA, is given in Supplementary Figures S1 and S2. Briefly, hyperspectral  datasets are subset to candidate endmember matrices by repeated simplex maximization in an increasing number of dimensions. Consequently, the same pixel/sample may be considered an endmember candidate in several such hyperdimensional spaces. The candidate endmember matrices are used as input to multiple semi-nonnegativity constrained matrix factorization models, a type of blind signal separation technique, where each model assumes a successively increasing number of source signals (Fig. \ref{fig:Ifigure1}b). Dimensionality is estimated from these blind signal separation models based on the information entropy (Shannon entropy \citep{Shannon1948}) of the reconstruction residuals and the reduction in sum of squared residuals for each model (error reduction). Individually these methods may be used with thresholding to estimate the dimensionality. However, the selection of threshold is often subjective. For both CD-DE and CD-EEA, information entropy and error reduction provides complimentary information on model performance based on metrics describing residual information in the reconstruction residuals (Shannon entropy) and a measure of model fit to the data (reduction in sum of squared residuals). 

\subsection{Benchmarking CD-DE  to dimensionality estimation techniques from remote sensing}

First we benchmarked the developed CD-DE to established methods in analytical chemistry and satellite remote sensing, hereinafter, collectively denoted Virtual Dimensionality methods; the Harsanyi-Farrand-Chang (HFC) method \citep{Chang2013} based on Neyman-Pearson detection theory, the Akaike Criterion (AIC) and Minimum Description Length (MDL)\citep{Wax2009} based on information theory, the Malinowski Factor Indicator Function (FIF) derived from factor analysis \citep{Malinowski1977}, and last, dimensionality estimated from the number of principal components required to explain at least \SI{95}{\percent} and \SI{99}{\percent} of the total variance, denoted $\lambda_{95}$ and $\lambda_{99}$ respectively. Initially we sought to benchmark CD-DE on its ability to handle datasets in which the number of samples/pixels $n$ is less than the number of channels/wavenumbers $p$, approximately equal $n \approx p$, and greater than the number of channels $n > p$. We generated pseudo-spectral datasets with a known number of endmembers, ranging from two to ten endmembers and $p = 1001$ channels, vector normalized to unit length, convexly mixed \textit{in silico} to $n = 300$, $1000$ and $5000$ samples and added white Gaussian noise to a signal-to-noise ratio (SNR) of 1000:1 to avoid strict linear dependencies (Table \ref{tab:ItableS1} in Methods). 

We found for $n = 300$ samples and $p = 1001$ channels, the developed CD-DE was the only method estimating the correct dimensionality (Supplementary Table S1). AIC, MDL and FIF were unable to accurately estimate the dimensionality for datasets where the number of samples was approximately equal or less than the number of channels (Supplementary Tables S1 and S2), while the explained variance thresholding measures, $\lambda_{95}$ and $\lambda_{99}$ generally underestimated the dimensionality for $n = 300$, $1000$ and $5000$, but was consistent in its estimations (Supplementary Tables S1, S2 and S3). In all three cases,  $n = 300$, $1000$, $5000$ the developed CD-DE method  correctly and consistently estimated the correct dimensionality based 30 repeats of CD-DE algorithm (Supplementary Tables S1, S2 and S3), as the only one of the tested methods (full confusion matrix in Supplementary Table S4). Based on Neyman-Pearson Detection Theory using a false-alarm probability of $P_F = 10^{-5}$, the HFC method was able to estimate the dimensionality for all three cases, but generally overestimating the dimensionality for $n = 300$ and $1000$ (Supplementary Tables S1 and S2). 

\begin{table}[t!]
	\small
	\centering
	\caption[Estimated dimensionalities for $n$~=~5000 spectra, SNR~=~10:1]{Comparison of estimated dimensionality for $n$~=~5000 spectra, SNR~=~10:1}
	\label{tab:Itable2}

	\begin{tabular}[c]{@{}p{1cm}lllllll@{}}
		\toprule
		Ground truth & CD$^{[a],[b]}$    & HFC$^{[b]}$   & AIC   & MDL   & FIF   & $\lambda_{95}$   & $\lambda_{99}$       \\ \midrule
		2 & 2 & 2 & 2 & 2 & 2 & 35 & 653 \\
		3 & 3(\SI{63}{\percent}) & 3 & 3 & 3 & 3 & 2 & 580 \\
		4 & 5 & 4 & 4 & 4 & 4 & 160 & 713 \\
		5 & 5 & 4 & 5 & 5 & 5 & 81 & 675 \\
		6 & 6 & 6 & 6 & 6 & 6 & 130 & 699 \\
		7 & 7 & 6 & 7 & 7 & 7 & 171 & 718 \\
		8 & 8 & 5 & 8 & 8 & 8 & 91 & 680 \\
		9 & 8 & 7 & 9 & 9 & 9 & 145 & 705 \\
		10 & 10 & 6 & 10 & 10 & 10 & 243 & 749 \\ \bottomrule
    
	\end{tabular}%

\raggedright
$^{[a]}$ Based on 30 repeats of the Chemical Dimensionality (CD) algorithm. For a full confusion matrix of the CD-algorithm see Supplementary Table S13. $^{[b]}$ Number in brackets denotes the percentage of 30 algorithm repeats which yielded the given estimation of dimensionality. $^{[c]}$ Using a false alarm probability of $10^{-5}$.
\end{table}

We then tested whether the accuracy of CD-DE was contingent on the size of the candidate endmember matrix. Using the mixture dataset containing $n = 5000, p = 1001$ at an SNR of 1000:1, we varied  the number of hyperdimensional simplex maximization repeats used to populate the candidate endmember matrix. We found no effect of varying size of the candidate endmember matrix by volume maximization of hyperdimensional simplexes up to a maximum of 14, 19 and 29 dimensions (Supplementary Table S5), translating to matrices of 119, 209 and 464 endmember candidates.

\subsection{Benchmarking CD-DE to Virtual Dimensionality methods with respect to signal noise}

We then examined the robustness of CD-DE in response to increasing noise levels, benchmarked to existing Virtual Dimensionality methods. Ground-true endmembers remained the same as in prior analysis, however, we generated four new noised datasets, each consisting of nine constituent datasets containing between two and ten endmembers, that were mixed by convex mixing \textit{in silico} to $n = 5000$ samples, and added white Gaussian noise to a signal-to-noise ratio of SNR =  100:1, 50:1, 20:1 and 10:1 respectively (see Methods).

CD-DE was unaffected by increasing noise levels for SNR=1000:1, 100:1, and 50:1 (Supplementary Tables S3, S6 and S7), correctly and consistently estimating the ground true number of endmembers across all 30 repeats of the algorithm confer confusion matrices in Supplementary Tables S4, S10 and S11. The same was true for AIC, MDL, FIF and HFC at all tested noise levels, however, HFC only yielded the ground true number of endmembers in four out of nine cases.  $\lambda_{95}$ and $\lambda_{99}$ were consistent, but generally underestimated the number of endmembers for SNR 100:1 and 50:1.

For SNR 20:1 the CD-DE estimated the correct dimensionality for all endmember datasets except the dataset containing three endmembers, and with minute variability for the dataset containing nine endmembers, for which 29 of 30 repeats estimated the correct dimensionality (Supplementary Table S8). At SNR 20:1 the $\lambda_{99}$ began to significantly overestimate the dimensionality, and for SNR 10:1 both $\lambda_{95}$ and $\lambda_{99}$ greatly overestimated the number of endmembers in all but one case (Table \ref{tab:Itable2}, and Supplementary Table S8). CD-DE remained generally consistent, only with variability in the dataset containing three endmembers, confer confusion matrices in Supplementary Tables S12 and S13. The estimated dimensionality was correct in 7 out of 9 cases, and never deviated more than one endmember from the ground truth (Table \ref{tab:Itable2}).

\subsection*{Benchmarking of CD-EEA endmember extraction}

\begin{figure*}[t!]
	\centering
	\includegraphics[width=0.9\textwidth]{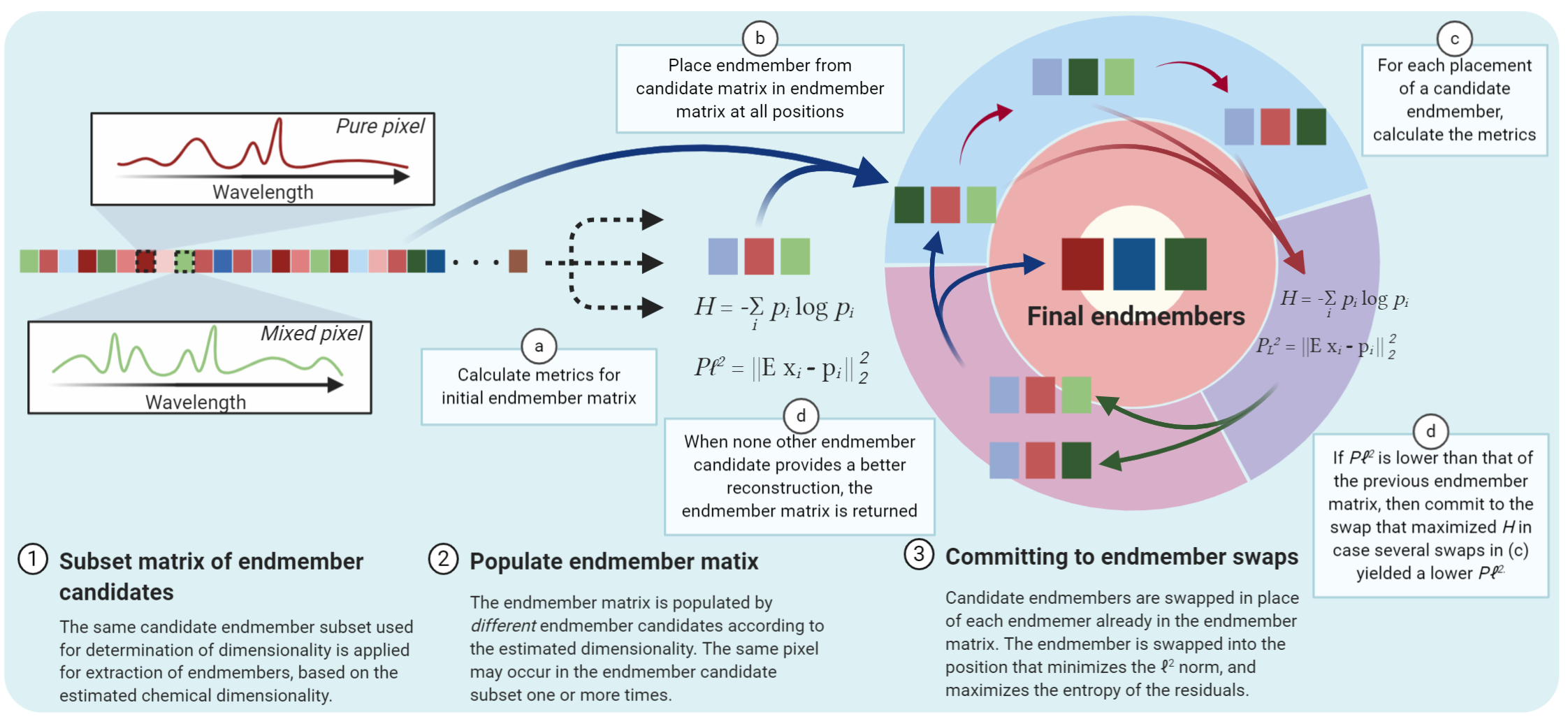}
	\caption[Comparison of endmembers extracted by CD-EEA and N-FINDR]{Schematic of CD-EEA endmember extraction algorithm. Final selection of endmembers is based on combined assessment of the squared $\ell^2$ norm of the residual and the Shannon entropy of the reconstruction residuals.}
	\label{fig:Ifigure2}
\end{figure*}

\begin{figure}[ht!]
	\centering
	\includegraphics[width=0.9\linewidth]{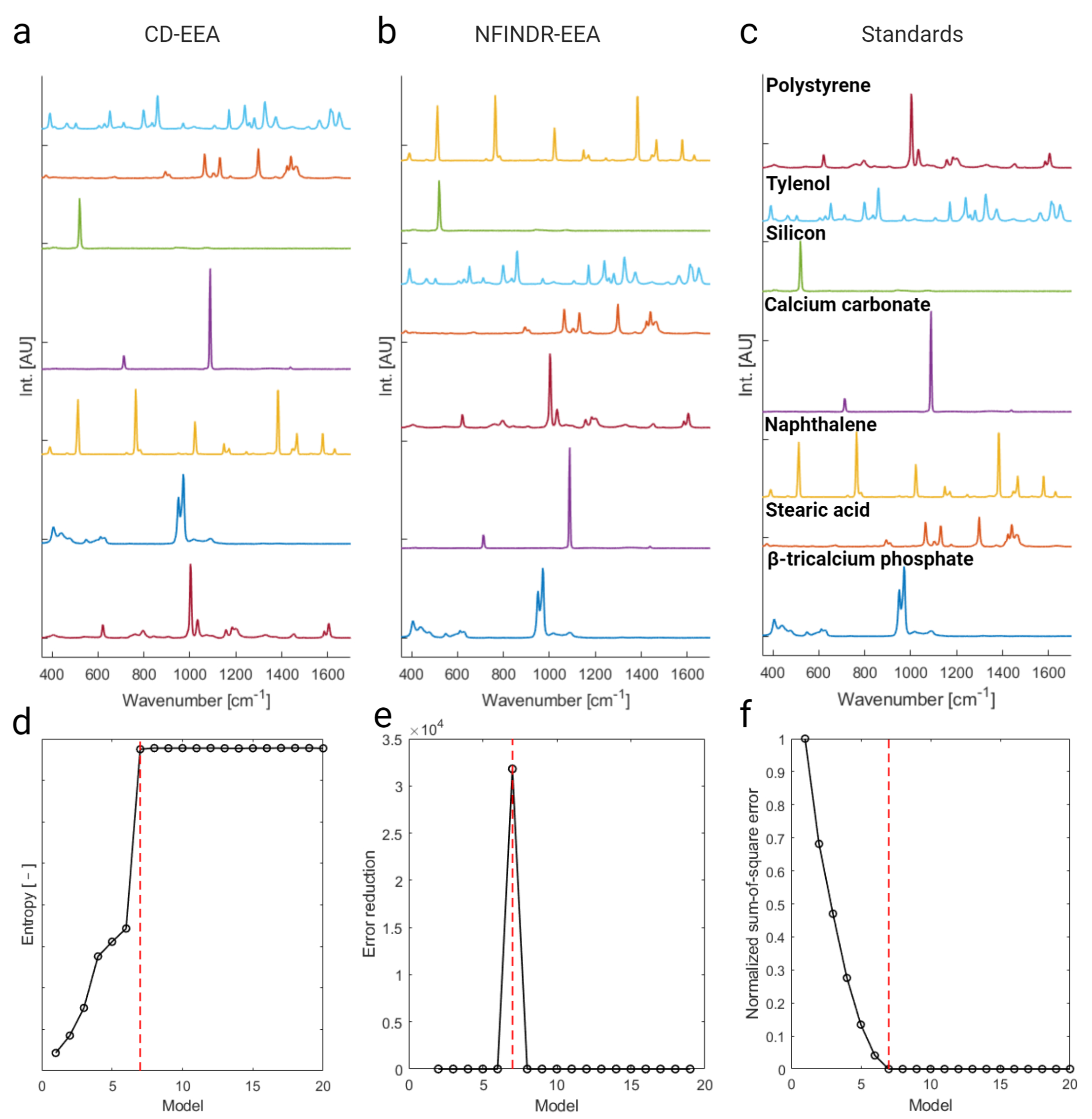}
	\caption[Comparison of endmembers extracted by CD-EEA and N-FINDR]{Endmember retrieval by \textbf{(a)} the CD-EEA endmember extraction algorithm and \textbf{(b)} the N-FINDR algorithm, with the ground true reference spectra in \textbf{(c)}. The measured spectra were baseline-corrected by a Whittaker filter cf. Supporting Information, mixed to 5000 samples and added white Gaussian noise to SNR = 1000:1. \textbf{(d)-(f)} model performance as a function of increasing the number of source signals for entropy \textbf{(d)}, error redution \textbf{(e)} and normalized sum-of-square residuals \textbf{(f)}. Red dashed line indicates the final estimation of chemical dimensionality ($k_{\text{CD}} = 7$) by the CD algorithm.  Line colors correspond to standards in (c).}
	\label{fig:Ifigure3}
\end{figure}

We then investigated if the developed framework could function as an endmember extraction algorithm (CD-EEA). The goal of an endmember extraction algorithm (EEA) is to retrieve the ``purest'' feature vector/spectrum for each distinct signal source, that combined describes the remaining mixture samples in the dataset. Using the dimensionality estimated by CD-DE, $k_{\text{CD}}$, CD-EEA extracts endmember spectra that 1) lowers the squared $\ell^2$ norm, and 2) if possible, maximizes the entropy of the reconstruction residuals (Fig. \ref{fig:Ifigure3}a, Methods). To evaluate the performance of the CD framework as an endmember extraction method, we initially benchmarked CD-EEA against N-FINDR, based on synthetic datasets containing 2, 5 and 10 endmembers vector normalized to unit length, and convexly mixed \textit{in silico} to 5000 samples with SNR = 1000:1. Out of the 5000 spectra, both CD-EEA and N-FINDR succeeded at recovering all ground true endmembers in all three cases (Supplementary Figures S3 to S5), suggesting that CD-EEA indeed can function as an endmember extraction algorithm.

To further substantiate the ability of CD-EEA as an endmember extraction technique, we collected Raman spectra of seven reference compounds; polystyrene, tylenol, silicon, calcium carbonate, naphthalene, stearic acid and \textbeta-tricalcium phosphate, varying especially by the complexity of their spectral signatures (Figure \ref{fig:Ifigure3}c). The seven measure references were mixed convexly \textit{in silico} to 5000 samples and added white Gaussian noise to SNR = 1000:1 to avoid strict linear dependencies. The Chemical Dimensionality algorithm correctly estimated $k_{\text{CD}} = 7$, as indicated by both the error reduction and associated increase in reconstruction entropy (Figure \ref{fig:Ifigure3}d-f). Although the CD-EEA and N-FINDR returns endmembers unordered, they both uncovered the ground true endmember spectra out of the 5000 mixture spectra (Figure \ref{fig:Ifigure3}a-b).

\begin{table}[t!]
	\small
	\centering
	\caption[Estimated dimensionality for experimental datasets]{Estimated dimensionality for experimental datasets}
	\label{tab:Itable3}
		\begin{tabular}[c]{@{}lp{1.1cm}p{1.2cm}lllllll@{}}
			\toprule
			Experimental dataset & Sample size & Number of channels & CD    & HFC$^{[a]}$   & AIC   & MDL  & FIF & $\lambda_{95}$   & $\lambda_{99}$ \\ \midrule
			Standards (Raman)$^{[b]}$ & 5000 & 1608 & 7 & 7 & 7 & 7 & 7 & 6 & 6 \\
			Mayonnaise (CARS)$^{[c]}$ & 287296 & 33 & 2 & 30 & 32 & 32 & 3 & 26 & 31 \\
			PLLA w. TGF-\textbeta3 (FTIR) & 836 & 216 & 5 & 2 & 215 & 97 & 99 & 2 & 5 \\
			SIS w. TGF-\textbeta3 (FTIR) & 848 & 751 & 4 & 2 & 209 & 107 & 104 & 2 & 8 \\  \bottomrule
			
		\end{tabular}%
		
\raggedright
$^{[a]}$ False alarm probability of $10^{-5}$. $^{[b]}$ Measured standards mixed \textit{in silico}. $^{[c]}$ Input data vector normalized to unit length prior estimation of dimensionality and extraction of endmembers.

\end{table}

\subsection{Combined CD-DE and CD-EEA image reconstructions for imaging oil-in-water emulsions by Coherent Anti-stokes Raman Scattering}

\begin{figure}[b!]
	\centering
	\includegraphics[width=0.8\linewidth]{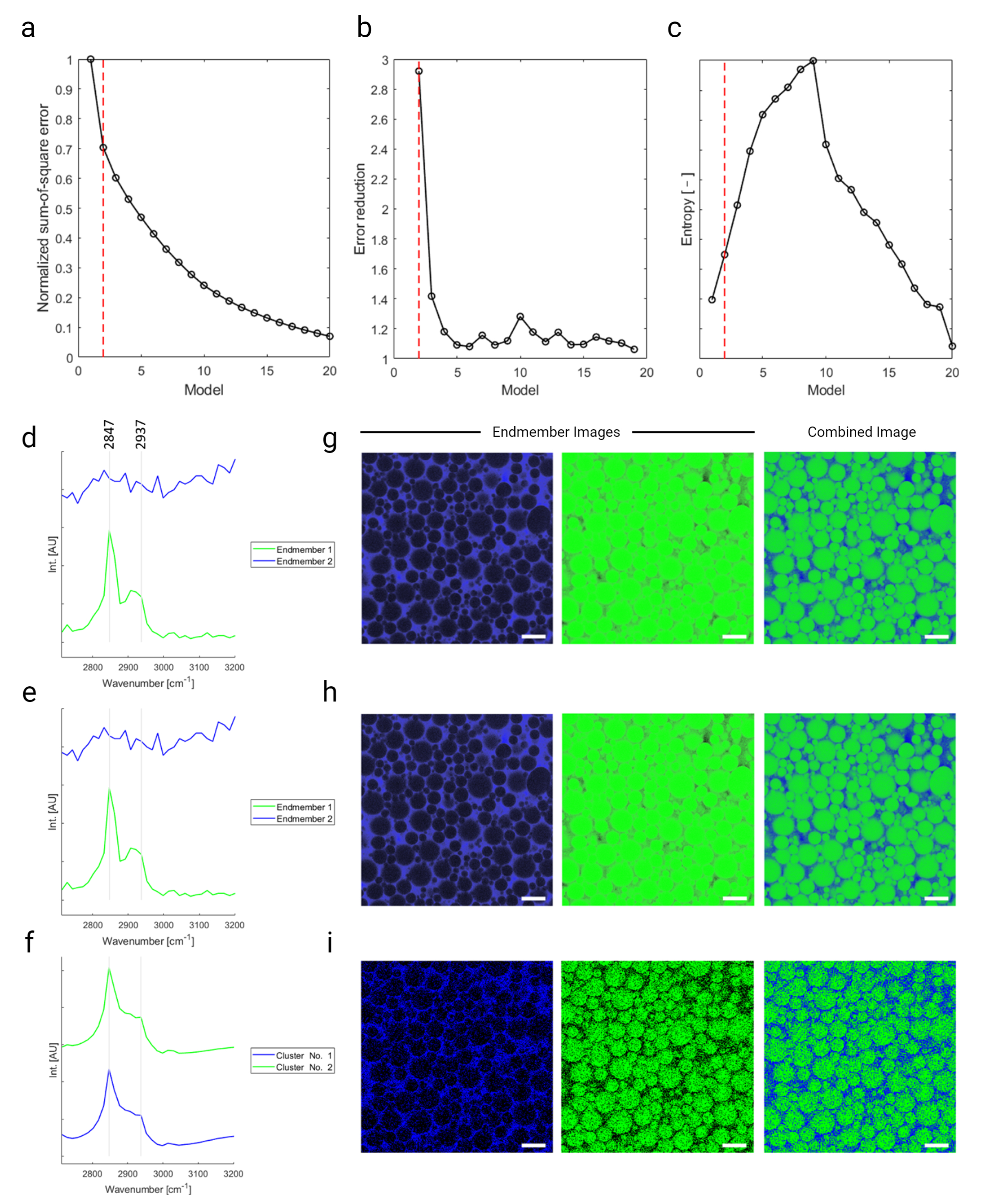}
	\caption[Diagnostic figures of CD-algorithm from CARS of mayonnaise]{\footnotesize Diagnostic figures from the CD-algorithm based on CARS microspectroscopy of mayonnaise, demonstrating \textbf{(a)} the max-normalized sum-of-square-error, \textbf{(b)} error reductions, of which a model containing 2 feature vectors was associated with the greatest error reduction, and \textbf{(c)} the reconstruction entropy for each of the 20 semi-nonnegativity constrained matrix factorization models. The reconstruction model containing two feature vectors was also associated with an increase in reconstruction entropy. Red dashed line is the dimensionality estimated by the CD algorithm ($k_{\text{CD}} = 2$). Endmember extraction and image reconstructions of immiscible phases in a mayonnaise oil-in-water emulsion. Endmembers extracted by \textbf{(d)} the CD-algorithm and \textbf{(e)} N-FINDR. \textbf{(f)} Cluster centroids from the pixels ascribed to their respective cluster (1 – blue, 2 – green). Image reconstructions are based on the extracted endmembers from \textbf{(g)} the CD endmember extraction algorithm (CD-EEA), and \textbf{(h)} N-FINDR. \textbf{(i)} cluster image from K-Means++ indicating the membership of a given pixel to either the blue or green cluster. Scale bar is \SI{10}{\micro\metre}. The color of the spectra in subfigures (d)-(f) corresponds to their image reconstructions in (g)-(i).}
	\label{fig:Ifigure4}
\end{figure}

Having benchmarked CD-DE and CD-EEA for estimation of data dimensionality and endmember extraction, we  sought to study more complex (bio)chemical system. We first studied liquid immiscibility in mayonnaise, a model oil-in-water emulsion, using Coherent Anti-stokes Raman Scattering (CARS). Contrary to most other microspectroscopy techniques, CARS signals are not easily used quantitatively owing to an inherent non-resonant background arising from coherent addition of off-resonant vibrational contributions, induced by the pump and Stokes fields \citep{Evans2008}, challenging both estimation of dimensionality and extraction of endmembers.

We compared endmembers extracted by CD-EEA to N-FINDR and cluster centroids from K-Means++ clustering, along with their associated image reconstructions, according the $k_{\text{CD}}$ estimated by CD-DE. For the CARS hyperspectral datacube of mayonnaise estimated dimensionality was $k_{\text{CD}} = 2$ (Figure \ref{fig:Ifigure4}a-c), a reasonable estimate given that mayonnaise is an oil-in-water emulsion. However, AIC, MDL, $\lambda_{95}$ and $\lambda_{99}$ all estimated 26 endmembers or more (Table \ref{tab:Itable3}).  

Remarkably, CD-EEA algorithm and N-FINDR both returned the same two endmembers given $k_{\text{CD}} = 2$ (Figure \ref{fig:Ifigure4}d-e). Endmember spectra were generally characterized by a background spectrum corresponding to the water-phase, and an endmember with peaks around \SI{2847}{\per\centi\metre} and \SI{2937}{\per\centi\metre}, assigned to C-H and \ce{CH3} stretching vibrations \citep{Cheng2007,LeicaMicrosystems2016}, consistent with the presence of lipid. The cluster centroids from K-Means++ indicates lipid in both endmembers, which may occur if pixels containing lipid is ascribed to both clusters (Figure \ref{fig:Ifigure3}f). Nonetheless, the K-Means++ algorithm succeeds at revealing the lipid-droplet microstructure (Figure \ref{fig:Ifigure4}i), as expected from an oil-in-water emulsion and previously demonstrated using both CARS and fluorescence microscopy \citep{Patil2019,Sorensen2019}. The same microstructure is observed in image reconstructions using endmembers returned by CD-EEA and N-FINDR (Figure \ref{fig:Ifigure3}g-h).

\subsection{CD-DE and CD-EEA for monitoring tissue-engineered extracellular matrices with FTIR microspectroscopy}

Label-free microspectroscopy plays a pivotal role in research and development of engineered tissue grown \textit{in vitro}. Monitoring cell growth, differentiation and migration in response to physical and mechanical cues served by the biomaterial scaffold is paramount for engineering personalized, biocompatible tissue replacements. Label-free microspectroscopy can offer a wealth of information with molecular specificity, however, the success hyperspectral imaging for monitoring is intimately related to the estimation of dimensionality and the ability of the endmember extraction algorithm to distinguish mixed spectral information from spectral information corresponding to an endmember. 

We engineered a scaffold from electrospun fibers of Poly-L-Lactic acid (PLLA), and a scaffold from Small Intestine Submucosa (SIS) derived from porcine jejunum. Both scaffolds were incubated with isolated primary bovine chondrocytes and transforming growth factor \textbeta3 (TGF-\textbeta3). After 4 weeks, scaffolds were imaged using FTIR microspectroscopy and analyzed using CD-DE and CD-EEA.

\begin{figure}[b!]
	\centering
	\includegraphics[width=0.6\linewidth]{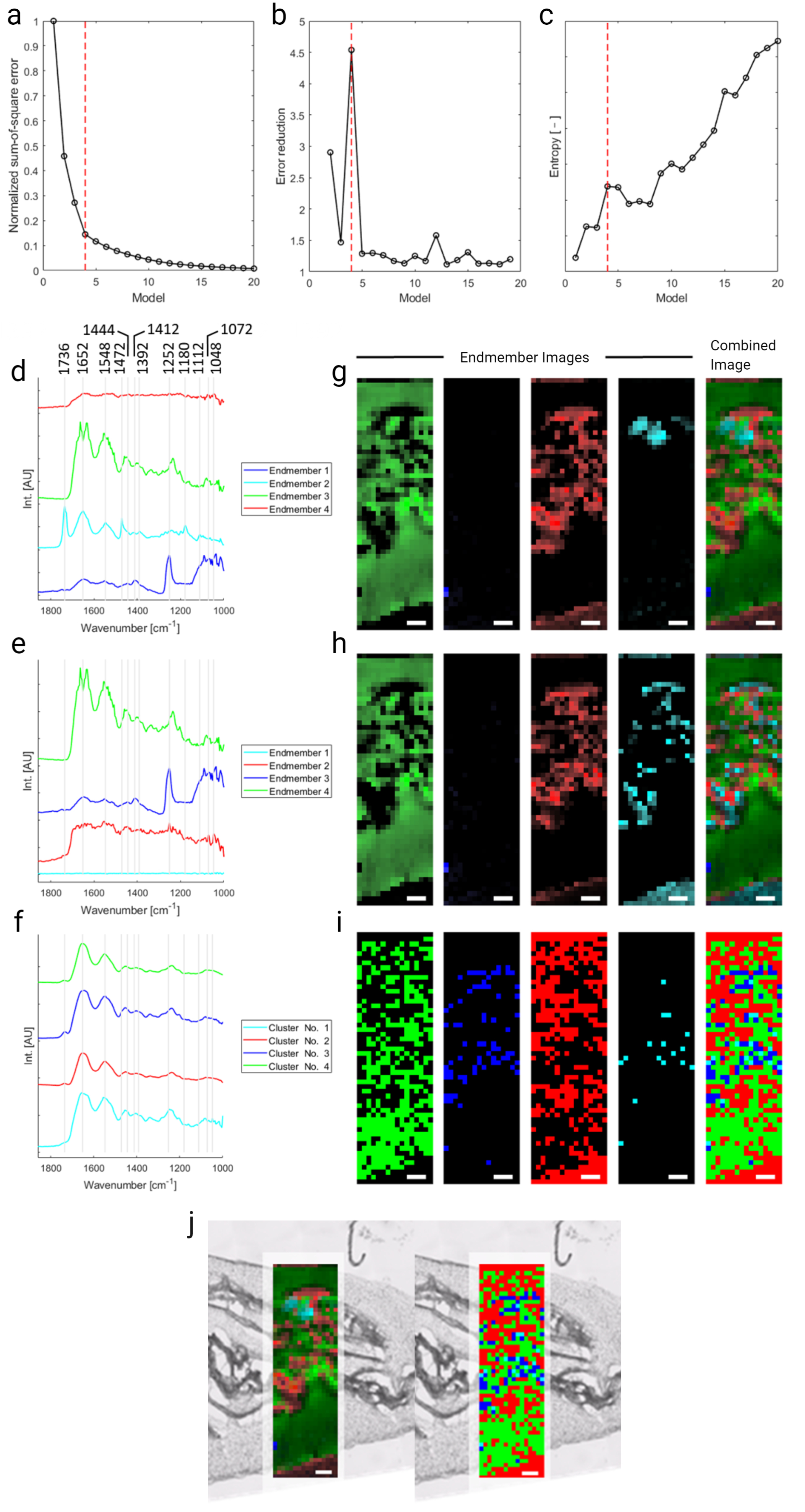}
	\caption[Diagnostic figures of CD-algorithm from FTIR of SIS scaffolds]{\footnotesize CD-DE Diagnostic figures based on FTIR microspectroscopy of SIS scaffold seeded with chondrocytes in presence of TGF-\textbeta3. \textbf{(a)} max-normalized sum-of-square-residuals, \textbf{(b)} corresponding error reduction and \textbf{(c)} reconstruction entropy for each of 20 semi-nonnegativity constrained matrix factorization models. Red dashed line is the dimensionality estimated by the algorithm ($k_{\text{CD}} = 4$). Endmember extraction and image reconstructions were obtained by  CD-EEA (\textbf{(d)} and \textbf{(g)}) and  N-FINDR (\textbf{(e)} and \textbf{(h)}). \textbf{(f)} shows cluster centroids and \textbf{(i)} cluster image from K-Means++. Number of endmembers extracted was determined by CD-DE. The color of the spectrum in subfigures (d)-(f) corresponds to image reconstructions in (g)-(i) of the same color. \textbf{(j)} Combined endmember image reconstructed from CD-EEA (left) and cluster membership image from K-Means++ clustering (right), overlaid on micrographs of the SIS scaffold. Scalebar is \SI{40}{\micro\metre}.}
	\label{fig:Ifigure6}
\end{figure}

The estimated chemical dimensionality for the PLLA and SIS scaffolds were $k_{\text{CD}} = 5$ and $k_{\text{CD}} = 4$ respectively (Table \ref{tab:Itable3}). Endmember extraction on the SIS-scaffold FTIR dataset yielded a similar endmember for the collagenous extracellular matrix as for the PLLA scaffolds, using both N-FINDR (endmember 4; green) and CD-EEA (endmember 3; green) for endmember extraction (Figure \ref{fig:Ifigure6}d-e). Absorptions at \SI{1652}{\per\centi\metre}, \SI{1548}{\per\centi\metre}, and \SI{1236}{\per\centi\metre} were assigned to amide I, amide II and amide III respectively, likewise displaying absorptions at \SI{1080}{\per\centi\metre} and \SI{1032}{\per\centi\metre}, that were assigned to C-O and C-O-C vibrations stemming from carbohydrate species in the tissue \citep{Belbachir2009}. Endmembers extracted are generally consistent between the two methods, however, CD-EEA discovers and extracts a spectrum containing spectral signatures at \SI{1736}{\per\centi\metre} and \SI{1472}{\per\centi\metre} (endmember 2; cyan; Figure \ref{fig:Ifigure6}d) in addition to the ECM-related amide signatures. The spectrum is not returned by N-FINDR (endmember 1; cyan; Figure \ref{fig:Ifigure6}e). The absorptions were assigned to carbonyl stretching (C=O; \SI{1736}{\per\centi\metre}) and methylene scissoring (C-H; \SI{1472}{\per\centi\metre}), suggesting the presence of lipid \citep{Mehrotra2000} in the submucosal layer of the small intestine. The associated endmember reconstructions indicate little co-localization between CD-EEA endmember 2 and the remaining endmembers (Figure \ref{fig:Ifigure6}g). Thus, this endmember resembles a lipid-rich region of the tissue. From the viewpoint of simplex volume maximization, N-FINDR returns a background spectrum (endmember 1; cyan; Figure \ref{fig:Ifigure6}b), confirmed by image reconstruction with the endmember in question (Figure \ref{fig:Ifigure6}h). Inspection of the cluster centroids from clustering with K-Means++ mainly indicates the presence of the three amide bands, confer findings with CD-EEA and N-FINDR, but are also indicative of weak signatures at \SI{1736}{\per\centi\metre} in all four endmembers (Figure \ref{fig:Ifigure6}f). The associated cluster membership images do not provide detailed microstructure of the tissue scaffold, and cluster centroids generally does not convey the molecular specificity offered by CD-EEA and N-FINDR (Figure \ref{fig:Ifigure6}d-e). More elaborate pre-processing would typically be required, such as background removal and extended multiplicative scattering corrections including spectral interference subtraction, to provide enhanced chemical and structural information. To visualize the contrast in structural and chemical information conveyed by the CD-EEA algorithm and K-Means++, we overlaid the combined endmember image reconstructed from endmembers extracted by CD-EEA (Figure \ref{fig:Ifigure6}g) and the combined cluster membership image from K-Means++ (Figure \ref{fig:Ifigure6}i), with micrographs of the SIS scaffold (Figure \ref{fig:Ifigure6}j). Results of hyperspectral analysis and associated image reconstructions of the PLLA scaffold can found in Supplementary Figures S6 and S7, including an overlay of the combined endmember reconstruction image from CD-EEA on the PLLA scaffold in Supplementary Figure S8.  

\section{Discussion}
\label{sec:Iconclusion}

We have developed an entropy-guided dimensionality estimation (CD-DE) and endmember extraction algorithm (CD-EEA) for (bio)chemical hyperspectral imagery. Our findings suggest a combination of simplex volume maximization, semi-nonnegativity constrained matrix factorization and information entropy, can both estimate the dimensionality of hyperspectral datasets across various types of microspectroscopy techniques and serve as basis for endmember extraction. To our knowledge, this is the first time that dimensionality estimation and endmember extraction is achieved using blind signal separation and Shannon entropy, across quantitative spectral information in spontaneous Raman and FTIR, as well as in CARS containing an inherent non-resonant background.

Estimated dimensionalities using the CD-DE were compared to established methods in satellite remote sensing, including the HFC, AIC, MDL, Malinowski FIF and canonical explained variance thresholding measures from analytical chemistry. CD-EEA endmember extraction for CARS, FTIR, spontaneous Raman and synthetic datasets was benchmarked to endmembers extracted by N-FINDR, and to cluster centroids from K-Means++ clustering of the data. In comparison to prior art \citep{Winter1999}, we relaxed the mixing requirements from convex- to conical mixing, and exploited relaxation of the non-negativity constraints on the self-modelling curve resolution models to account for weak negativity in intensities, for instance occurring as a part of the preprocessing of the hyperspectral datasets with baseline removal algorithms. We especially noted that although all Virtual Dimensionality methods were able to estimate dimensionalities of all experimental datasets, the estimates were for certain experimental datasets unreasonably high, predicting the presence of more than 100 endmembers in a single hyperspectral dataset (Table \ref{tab:Itable3}). Consequently, we argue that CD-DE is less dependent on i) the type of imaging/spectrometry and ii) the dimensions of the unfolded hyperspectral datamatrix, i.e. the number of samples and channels available in the datacube, than other Virtual Dimensionality methods.

The generic applicability offered by entropy-guided estimation of dimensionality and endmember extraction, as demonstrated by the plurality of samples and hyperspectral imaging methods used for demonstrating algorithm, serves as illustration of how CD-DE and CD-EEA  can prove to be a powerful for interrogating chemical and biochemical systems, from gastrophysics to tissue engineering, especially by consolidating hyperspectral molecular information with sample microstructure.

\section{Acknowledgements}
We are grateful to the Danish Molecular Biomedical Imaging Center (DaMBIC) at the University of Southern Denmark, for access to the CARS microscope. S. Vilms Pedersen gratefully acknowledges the Danish Ministry of Higher Education and Science for funding mobility in the duration of the project, under the EliteForsk Travel Fellowship. E. C. Arnspang was funded by the Villum Young Investigator grant (Project 19105). A. Runge Walther is supported by the Independent Research Fund Denmark (DFF-7017-00163). A. Callanan thanks the Irish Research Council for Science, Engineering and Technology (IRCSET)–Marie Curie International Mobility Fellowship co-funded grant PD/2010/INSP/1948 for funding. We would also like to thank Particle3D Aps for kind donation of \textbeta-tricalcium phosphate, and Mathias Porsmose Clausen from SDU FoodLab for kind donation of the mayonnaise. Select figures in this paper were drawn using BioRender.

\newpage
\appendix
\renewcommand\thefigure{S\arabic{figure}} 
\renewcommand\thetable{S\arabic{table}}
\setcounter{figure}{0}
\setcounter{table}{0}
\section*{Supporting Information}
\subsection*{Supplementary tables}
The CD-DE algorithm was compared performance-wise to established Virtual Dimensionality methods, under scenarios where 1) the sample size of the input datamatrix varied from 300-5000 samples (SNR = 1000:1), and 2) signal-to-noise ratios from 1000:1 to 10:1 for $n$ = 5000 samples. For the CD-DE algorithm alone, was the effect of varying the simplex maximization repeats ($g$) also tested, i.e. the number of samples carried over from the input datamatrix to the reduced candidate endmember matrix \textbf{V}. All benchmarking was carried out on the synthesized datasets containing a known number of endmembers, from two to ten endmembers, sufficient for most uses of microspectroscopy.

\subsection*{FTIR microspectroscopy of PLLA scaffold}
Using the PLLA and SIS scaffolds incubated with isolated primary bovine chondrocytes and TGF-\textbeta3, we estimated the Chemical Dimensionality using the CD-algorithm and observed the endmembers returned from CD-EEA and N-FINDR, as well as the cluster centroids from K-Means++. The estimated chemical dimensionality for the PLLA and SIS scaffolds were $k_{\text{CD}} = 5$ and $k_{\text{CD}} = 4$ respectively. 

While the maximum error reduction coincided with an increase in entropy for the SIS scaffold (Figure 5b-c in the paper), this was not the case for the PLLA scaffold. Here, the semi-nonnegativity constrained matrix factorization models containing between two and four feature vectors were associated with consecutively decreasing reconstruction entropies, despite the maximum error reduction was observed for the model containing two feature vectors. 

An increase in entropy was first observed for $k_{\text{CD}} = 5$, that noteworthily, also resulted in a peak, albeit smaller compared to the maximum, in the error reduction (Figure \ref{fig:IfigureS4} in Supplementary Information). This raised the question whether there were grounds for estimating a higher dimensionality based on the reconstruction entropy. Indeed, if two endmembers were extracted to represent a single endmember species \textit{de facto}, one would expect close co-localization  between the two endmembers. However, as observed in Figure \ref{fig:IfigureS5}d, the brightest pixels of endmember 5 (cyan), are spatially exclusive to the pixels associating with endmember 4 (magenta). Inspecting the corresponding endmember spectra reveals spectral differences in the low-wavenumber region from \SIrange{1100}{1000}{\per\centi\metre} (Figure \ref{fig:IfigureS5}a) across endmembers 2 (red), 4 (magenta) and 5 (cyan), assigned to C–O stretching vibration in PLLA \citep{Chieng2014}  and glycosaminoglycans \citep{Spalazzi2013}. Nonetheless, the remaining spectral features corresponding to PLLA (\SI{1756}{\per\centi\metre}, C=O; \SI{1452}{\per\centi\metre}, \SI{1384}{\per\centi\metre}, \SI{1364}{\per\centi\metre}, C-H) and the extracellular matrix (ECM; \SI{1652}{\per\centi\metre}, amide I; \SI{1548}{\per\centi\metre}, amide II; \SI{1236}{\per\centi\metre}, amide III) are present in each of the three endmembers (Figure \ref{fig:IfigureS5}a). Taken together, these observations suggest the algorithm is sensitive toward narrow-band spectral dissimilarity, even if features in remaining parts of the spectrum correlate to other endmembers.

\newpage

\section*{Supplementary figures}
\begin{figure}[ht!]
	\centering
	\includegraphics[width=11.4cm]{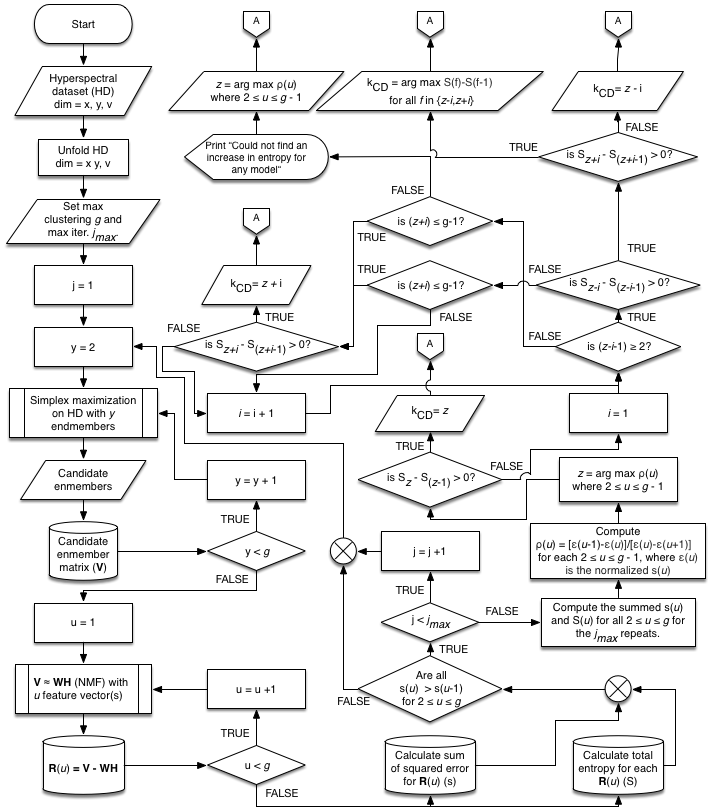}
	\caption[Algorithm flowchart for estimating the Chemical Dimensionality]{Algorithm flowchart for estimating the Chemical Dimensionality. Off-page connector labelled ‘A’ directs to the endmember extraction sub-algorithm.}
	\label{fig:IschematicS1}
\end{figure}

\begin{figure}[]
	\centering
	\includegraphics[width=11.4cm]{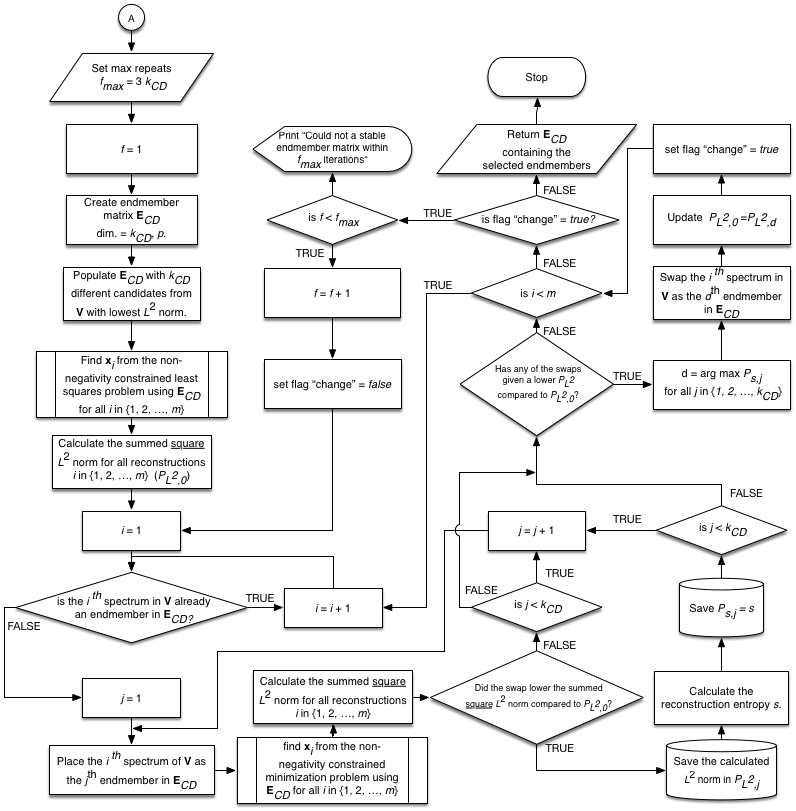}
	\caption[Algorithm flowchart for extraction of endmembers]{Algorithm flowchart for extraction of endmembers by minimization of the $L^2$ norm and subsequent entropy maximization.}
	\label{fig:IschematicS2}
\end{figure}

\begin{figure}[h!]
	\centering
	\includegraphics[width=0.85\linewidth]{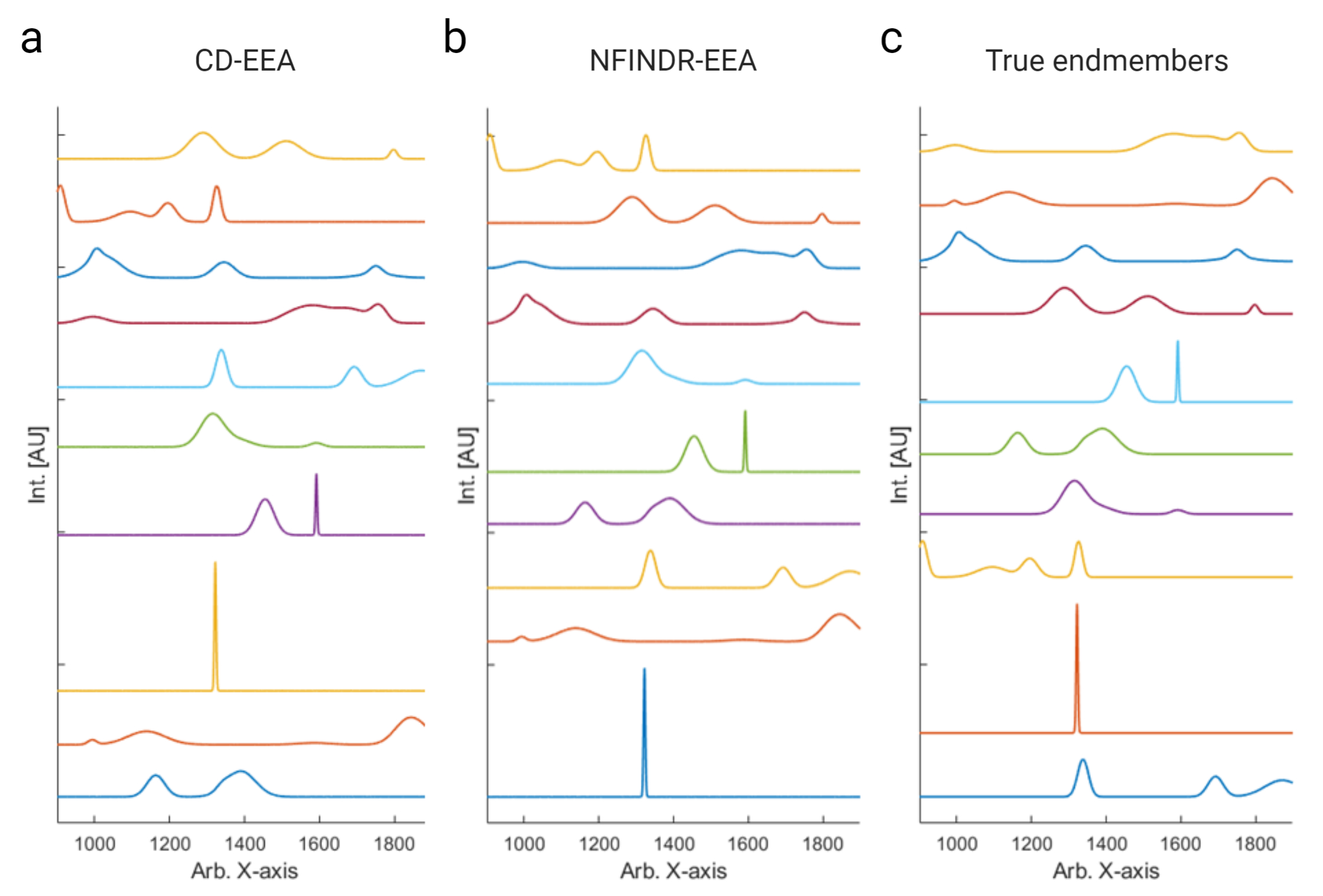}
	\caption[Ten endmembers extracted by CD-EEA and N-FINDR compared to ground true endmembers]{Endmembers extracted by \textbf{(a)} the Chemical Dimensionality Endmember Extraction algorithm (CD-EEA), \textbf{(b)} N-FINDR, compared to \textbf{(c)} the ground true endmembers. Ten ground true endmembers were convexly mixed to 5000 samples and added white Gaussian noise to SNR = 1000:1. Owing to random initialization in CD-EEA and N-FINDR, endmembers are returned unordered.}
	\label{fig:IfigureS1}
\end{figure}
\begin{figure}[h!]
	\centering
	\includegraphics[width=0.85\linewidth]{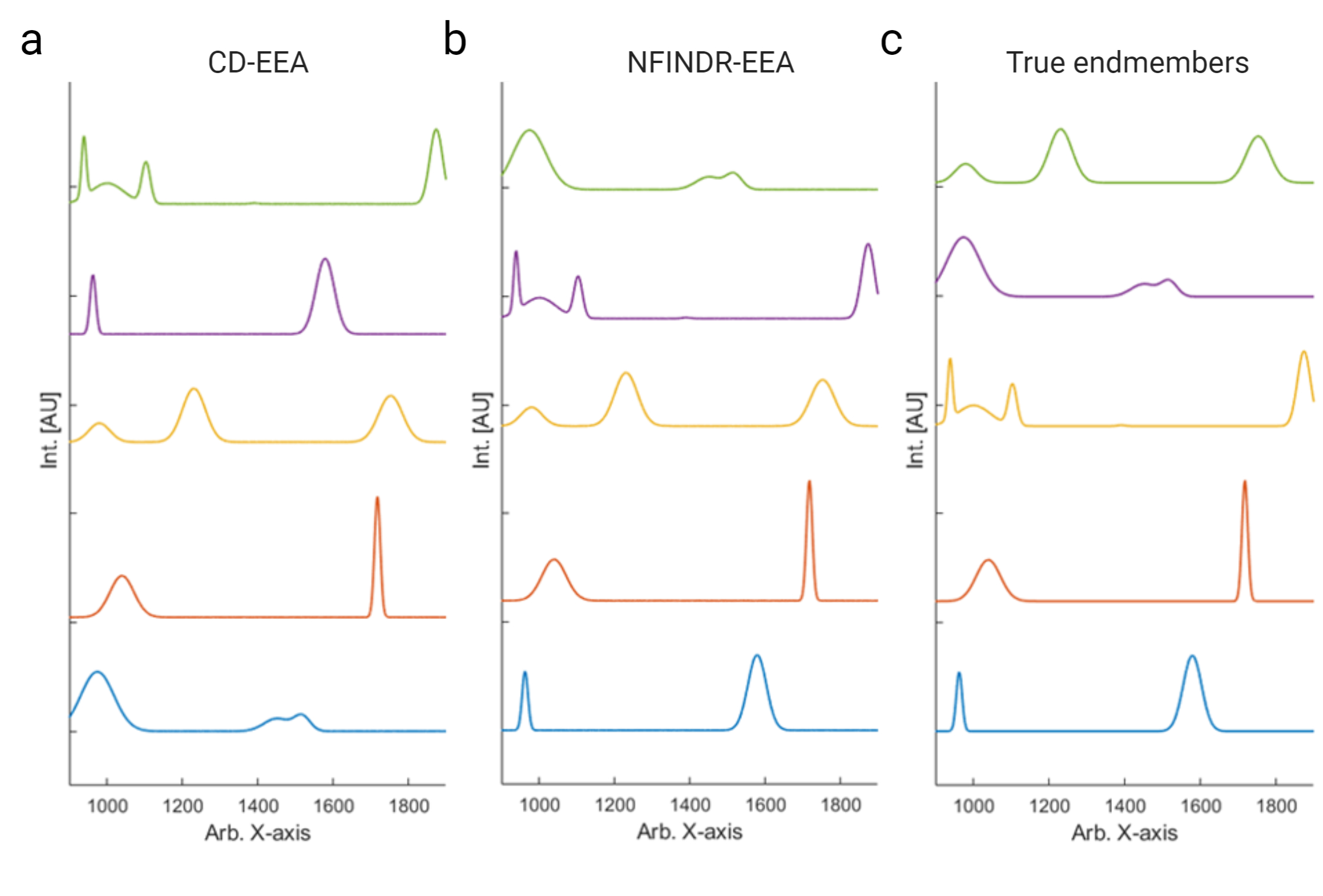}
	\caption[Five endmembers extracted by CD-EEA and N-FINDR compared to ground true endmembers]{Endmembers extracted by \textbf{(a)} the Chemical Dimensionality Endmember Extraction algorithm (CD-EEA), \textbf{(b)} N-FINDR, compared to \textbf{(c)} the ground true endmembers. Five ground true endmembers were convexly mixed to 5000 samples and added white Gaussian noise to SNR = 1000:1. Owing to random initialization in CD-EEA and N-FINDR, endmembers are returned unordered.}
	\label{fig:IfigureS2}
\end{figure}

\newpage

\begin{figure}[ht!]
	\centering
	\includegraphics[width=0.85\linewidth]{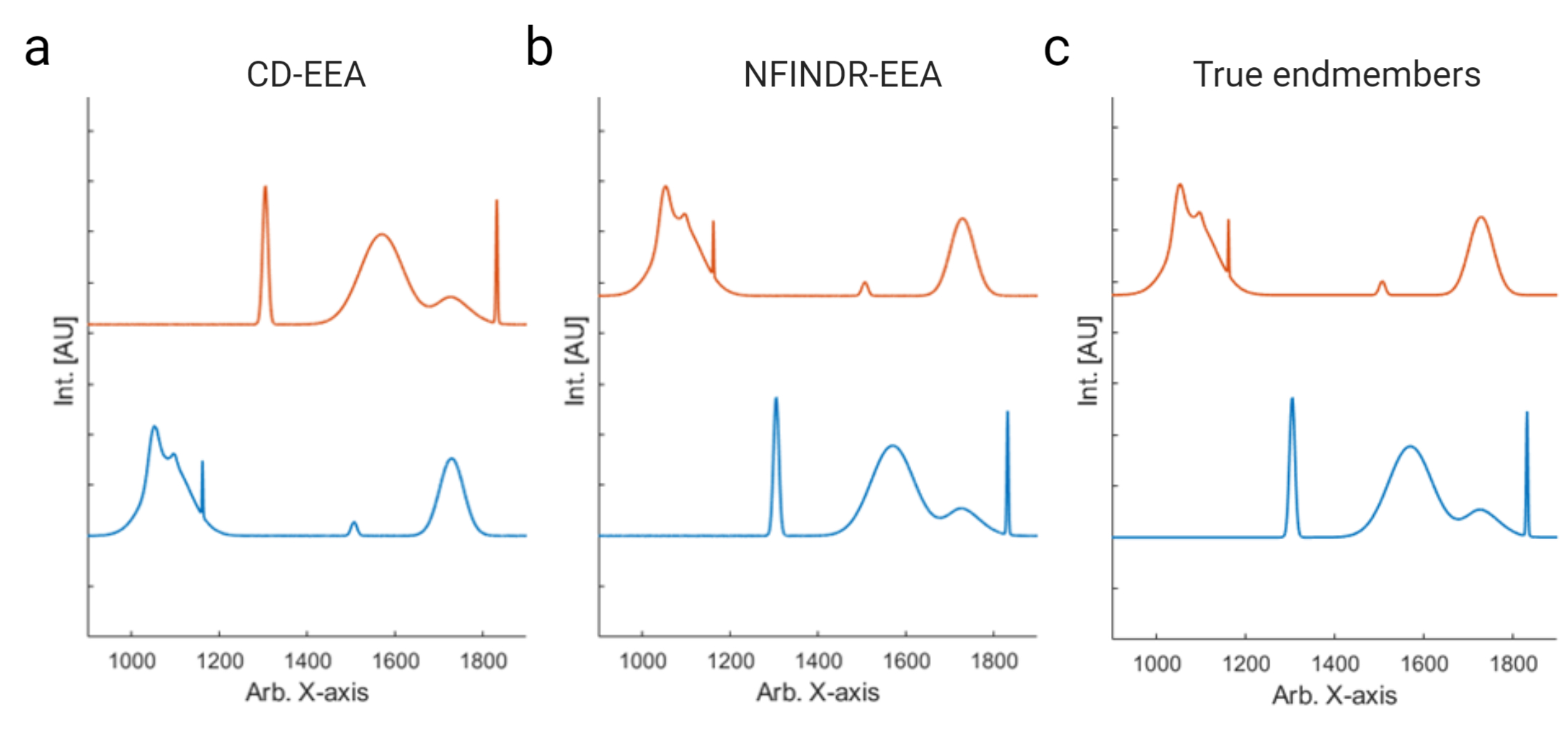}
	\caption[Three endmembers extracted by CD-EEA and N-FINDR compared to ground true endmembers]{Endmembers extracted by \textbf{(a)} the Chemical Dimensionality Endmember Extraction algorithm (CD-EEA), \textbf{(b)} N-FINDR, compared to \textbf{(c)} the ground true endmembers. Two ground true endmembers were convexly mixed to 5000 samples and added white Gaussian noise to SNR = 1000:1. Owing to random initialization in CD-EEA and N-FINDR, endmembers are returned unordered.}
	\label{fig:IfigureS3}
\end{figure}

\begin{figure}[ht!]
	\centering
	\includegraphics[width=0.9\linewidth]{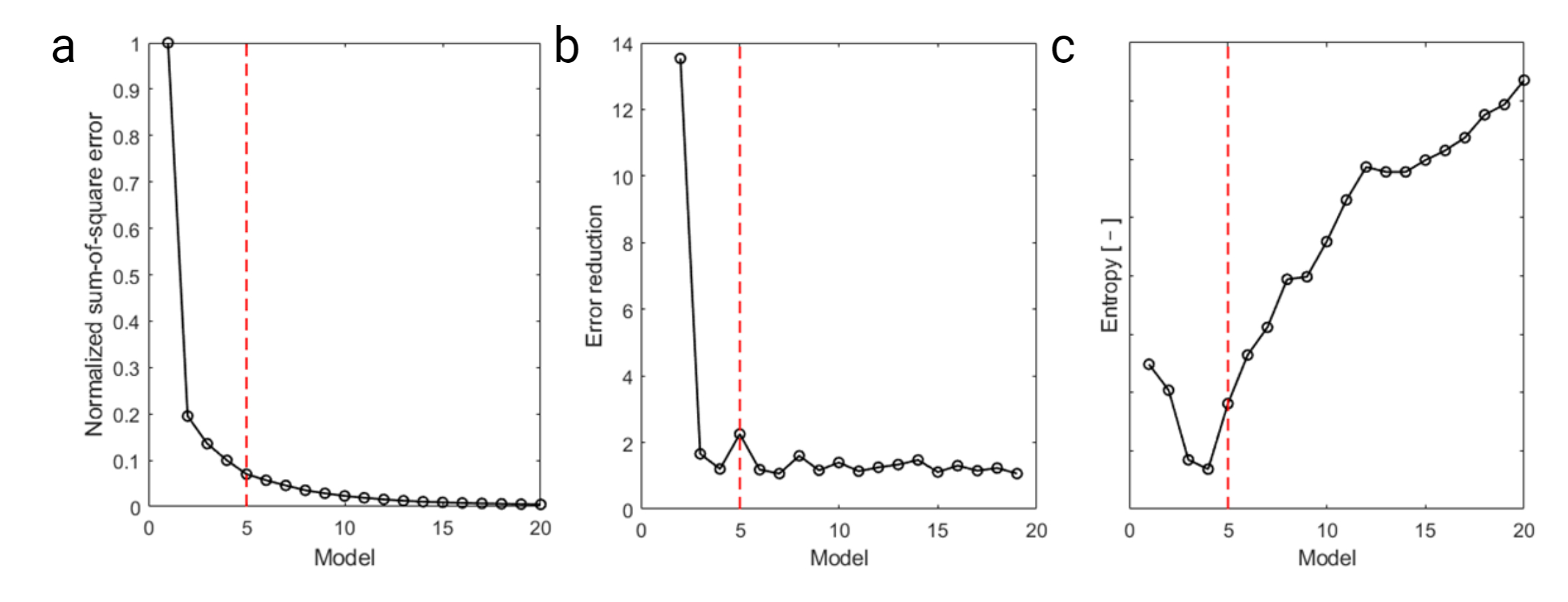}
	\caption[CD-algorithm diagnostic figures from FTIR of PLLA scaffold]{Diagnostic figures from the CD-algorithm based on FTIR microspectroscopy of PLLA scaffold seeded with chondrocytes in presence of TGF-\textbeta3. \textbf{(a)} max-normalized sum-of-square-error for each of 20 semi-nonnegativity constrained matrix factorization models, \textbf{(b)} the corresponding error reduction and \textbf{(c)} reconstruction entropy for each of the 20 semi-NMF models. Red dashed line is the dimensionality estimated by the algorithm ($k_{\text{CD}}$~=~5).}
	\label{fig:IfigureS4}
\end{figure}

\begin{figure}[h!]
	\centering
	\includegraphics[width=1\linewidth]{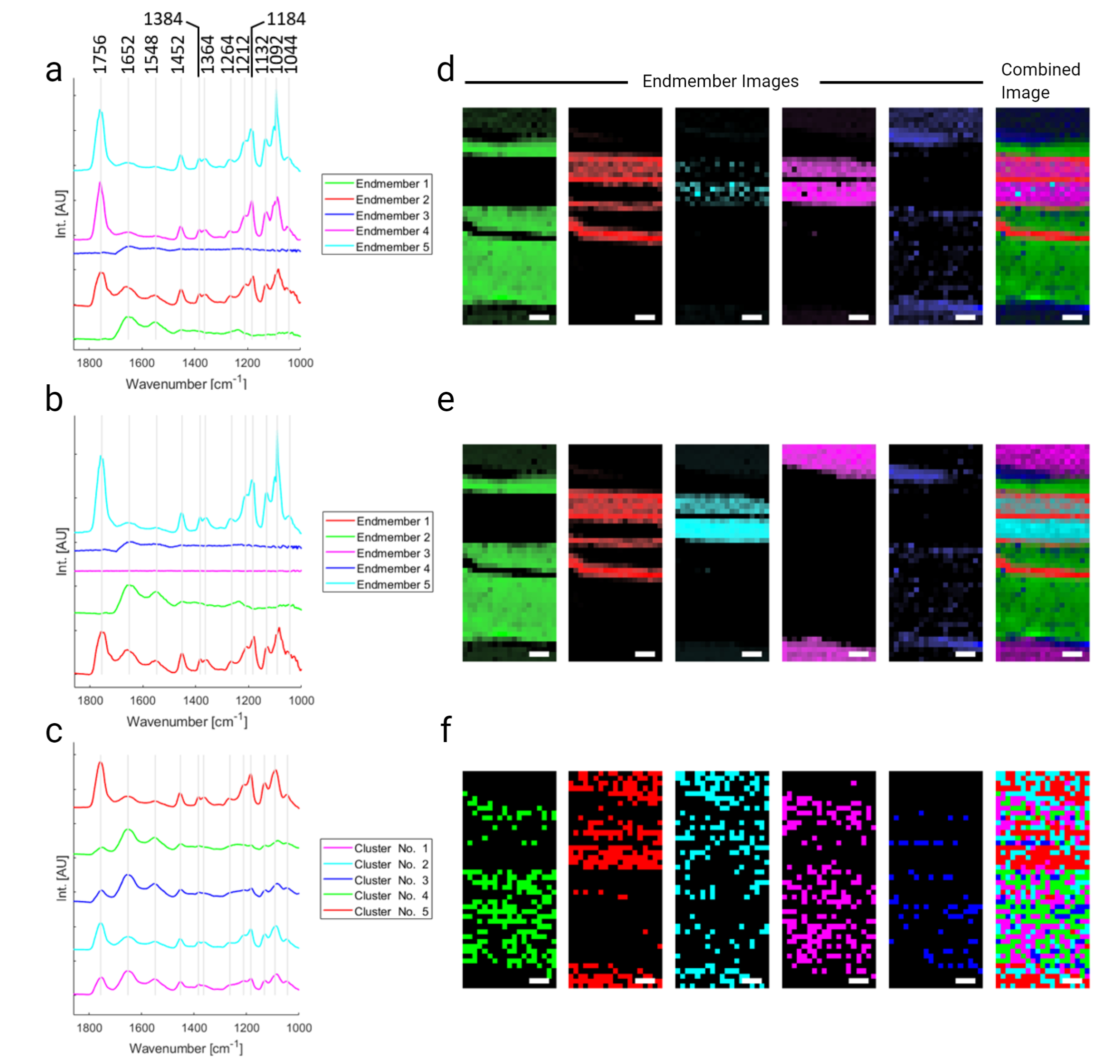}
	\caption[FTIR microspectroscopy, endmember extraction and image reconstruction of PLLA scaffold]{FTIR microspectroscopy, endmember extraction and image reconstructions of PLLA scaffold. Endmember extraction and image reconstructions were obtained by \textbf{(a)} and \textbf{(d)} the Chemical Dimensionality algorithm, \textbf{(b)} and \textbf{(e)} N-FINDR simplex maximization algorithm. \textbf{(c)} shows cluster centroids and \textbf{(f)} cluster image from K-Means++. Scale bar is \SI{40}{\micro\metre}. Number of endmembers extracted was determined by the Chemical Dimensionality algorithm. The color of the spectrum in subfigures (a)-(c) corresponds to image reconstructions in (d)-(f) of the same color.}
	\label{fig:IfigureS5}
\end{figure}

\begin{figure}[h!]
	\centering
	\includegraphics[width=0.75\linewidth]{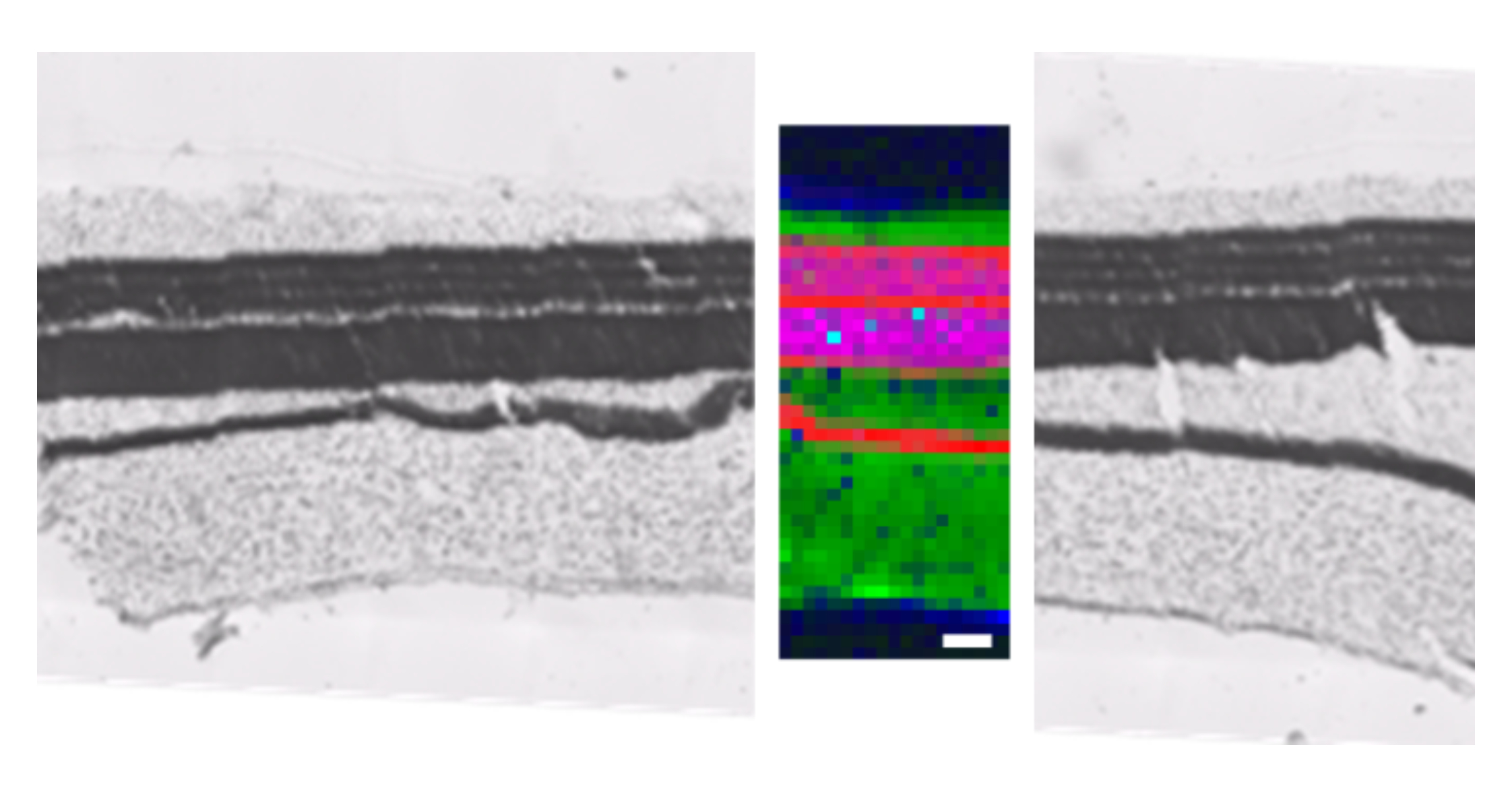}
	\caption[Overlaid PLLA scaffold micrograph with CD-EEA reconstructed image]{Combined endmember image reconstructed from CD-EEA overlaid on micrograph of the PLLA scaffold. Scale bar is \SI{40}{\micro\metre}.}
	\label{fig:IfigureS6}
\end{figure}

\FloatBarrier
\section*{Supplementary tables}
\begin{table}[ht!]
	\small
	\renewcommand{\arraystretch}{1.5}%
	\centering
	\caption[Estimated dimensionalities for $n$~=~300 spectra, SNR~=~1000:1]{Comparison of estimated dimensionality for $n$~=~300 spectra, SNR~=~1000:1}
	\label{tab:ItableS0}
	\begin{tabular}{@{}lp{1.3cm}p{1.3cm}p{1.3cm}p{1.3cm}p{1.3cm}p{1.3cm}p{1.3cm}@{}}
		\toprule
		Ground truth & CD$^{[a]}$    & HFC$^{[b]}$   & AIC   & MDL   & FIF   & $\lambda_{95}$   & $\lambda_{99}$       \\ \midrule
		2 & 2 & 5 & -- & -- & -- & 1 & 1 \\
		3 & 3 & 13 & -- & --& -- & 2 & 2 \\
		4 & 4 & 9 & -- & -- & -- & 3 & 3 \\
		5 & 5 & 10 & -- & -- & -- & 4 & 4 \\
		6 & 6 & 18 & -- & -- & -- & 4 & 5 \\
		7 & 7 & 15 & -- & -- & -- & 6 & 6 \\
		8 & 8 & 17 & -- & -- & -- & 6 & 7 \\
		9 & 9 & 16 & -- & -- & -- & 7 & 8 \\
		10 & 10 & 25 & -- & -- & -- & 8 & 9 \\ \bottomrule
		\multicolumn{8}{p{11.7cm}}{\footnotesize \begin{tabular}[c]{@{}p{\linewidth}@{}}$^{[a]}$ Based on 30 repeats of the Chemical Dimensionality (CD) algorithm. Full confusion matrix in Table \ref{tab:ItableS4}.  \\ $^{[b]}$ False alarm probability of $10^{-5}$.  \end{tabular}}     
	\end{tabular}
\end{table}

\begin{table}[h!]
	\small
	\renewcommand{\arraystretch}{1.5}%
	\centering
	\caption[Estimated dimensionalities for $n$~=~1000 spectra, SNR~=~1000:1]{Comparison of estimated dimensionality for $n$~=~1000 spectra, SNR~=~1000:1}
	\label{tab:ItableS2}
	\begin{tabular}{@{}lp{1.3cm}p{1.3cm}p{1.3cm}p{1.3cm}p{1.3cm}p{1.3cm}p{1.3cm}@{}}
		\toprule
		Ground truth & CD$^{[a]}$    & HFC$^{[b]}$   & AIC   & MDL   & FIF   & $\lambda_{95}$   & $\lambda_{99}$       \\ \midrule
		2 & 2 & 6 & -- & -- & -- & 1 & 1 \\
		3 & 3 & 8 & 1000 & 3 & 1000 & 2 & 2 \\
		4 & 4 & 9 & --& -- & -- & 3 & 3 \\
		5 & 5 & 9 & -- & -- & -- & 4 & 4 \\
		6 & 6 & 11 & 1000 & 6 & 1000 & 4 & 5 \\
		7 & 7 & 11 & 1000 & 7 & 1000 & 6 & 6 \\
		8 & 8 & 7 & -- & -- & -- & 6 & 7 \\
		9 & 9 & 9 & 1000 & 9 & 1000 & 7 & 8 \\
		10 & 10 & 8 & 1000 & 10 & 1000 & 8 & 9 \\ \bottomrule
		\multicolumn{8}{p{11.7cm}}{\footnotesize \begin{tabular}[c]{@{}p{\linewidth}@{}}$^{[a]}$ Based on 30 repeats of the Chemical Dimensionality algorithm. Confusion matrix in Table \ref{tab:ItableS4}.  \\ $^{[b]}$ False alarm probability of $10^{-5}$.  \end{tabular}}     
	\end{tabular}
\end{table}

\begin{table}[h!]
	\small
	\renewcommand{\arraystretch}{1.5}%
	\centering
	\caption[Estimated dimensionalities for $n$~=~5000 spectra, SNR~=~1000:1]{Comparison of estimated dimensionality for $n$~=~5000 spectra, SNR~=~1000:1}
	\label{tab:ItableS3}
	\begin{tabular}{@{}lp{1.3cm}p{1.3cm}p{1.3cm}p{1.3cm}p{1.3cm}p{1.3cm}p{1.3cm}@{}}
		\toprule
		Ground truth & CD$^{[a]}$    & HFC$^{[b]}$   & AIC   & MDL   & FIF   & $\lambda_{95}$   & $\lambda_{99}$       \\ \midrule
		2 & 2 & 2 & 2 & 2 & 2 & 1 & 1 \\
		3 & 3 & 3 & 3 & 3 & 3 & 2 & 2 \\
		4 & 4 & 4 & 4 & 4 & 4 & 3 & 3 \\
		5 & 5 & 4 & 5 & 5 & 5 & 4 & 4 \\
		6 & 6 & 6 & 6 & 6 & 6 & 4 & 5 \\
		7 & 7 & 6 & 7 & 7 & 7 & 6 & 6 \\
		8 & 8 & 5 & 8 & 8 & 8 & 6 & 7 \\
		9 & 9 & 7 & 9 & 9 & 9 & 7 & 8 \\
		10 & 10 & 6 & 10 & 10 & 10 & 8 & 9 \\ \bottomrule
		\multicolumn{8}{p{11.7cm}}{\footnotesize \begin{tabular}[c]{@{}p{\linewidth}@{}}$^{[a]}$ Based on 30 repeats of the Chemical Dimensionality algorithm. Confusion matrix in Table \ref{tab:ItableS4}.  \\ $^{[b]}$ False alarm probability of $10^{-5}$.  \end{tabular}}     
	\end{tabular}
\end{table}

\begin{table}[]
	\renewcommand{\arraystretch}{1.5}%
	\centering
	\caption[CD-algorithm confusion matrix for 300/1000/5000 samples, SNR = 1000:1]{Confusion matrix of CD-algorithm for 30 repeats, with endmembers mixed to 300/1000/5000 spectra, SNR~=~1000:1}
	\label{tab:ItableS4}
	\resizebox{\textwidth}{!}{%
		\begin{tabular}{@{}lccccccccc@{}}
			\toprule
			& \multicolumn{9}{c}{Chemical Dimensionality Algorithm} \\ \cmidrule(r){2-10}
			Ground truth & 2    & 3   & 4   & 5   & 6   & 7   & 8   & 9   & 10   \\ \midrule
			2            & 30/30/30     & 0    & 0    & 0    & 0    & 0    & 0    & 0    & 0     \\
			3            & 0     & 30/30/30    & 0    & 0    & 0    & 0    & 0    & 0    & 0     \\
			4            & 0     & 0    & 30/30/30    & 0    & 0    & 0    & 0    & 0    & 0     \\
			5            & 0     & 0    & 0    & 30/30/30    & 0    & 0    & 0    & 0    & 0     \\
			6            & 0     & 0    & 0    & 0    & 30/30/30    & 0    & 0    & 0    & 0     \\
			7            & 0     & 0    & 0    & 0    & 0    & 30/30/30    & 0    & 0    & 0     \\
			8            & 0     & 0    & 0    & 0    & 0    & 0    & 30/30/30    & 0    & 0     \\
			9            & 0     & 0    & 0    & 0    & 0    & 0    & 0    & 30/30/30    & 0     \\
			10           & 0     & 0    & 0    & 0    & 0    & 0    & 0    & 0    & 30/30/30     \\ \bottomrule
		\end{tabular}%
	}
\end{table}

\begin{table}[]
	\renewcommand{\arraystretch}{1.5}%
	\centering
	\caption[CD-algorithm confusion matrix for $n$ = 5000 with 15/20/30 simplex maximization repeats]{Confusion matrix of CD-algorithm for 30 repeats, with endmembers mixed to $n$~=~5000 spectra, SNR~=~1000:1 for 15/20/30 simplex maximization repeats}
	\label{tab:ItableS13}
	\resizebox{\textwidth}{!}{%
		\begin{tabular}{@{}lccccccccc@{}}
			\toprule
			& \multicolumn{9}{c}{Chemical Dimensionality Algorithm} \\ \cmidrule(r){2-10}
			Ground truth & 2    & 3   & 4   & 5   & 6   & 7   & 8   & 9   & 10   \\ \midrule
			2            & 30/30/30     & 0    & 0    & 0    & 0    & 0    & 0    & 0    & 0     \\
			3            & 0     & 30/30/30    & 0    & 0    & 0    & 0    & 0    & 0    & 0     \\
			4            & 0     & 0    & 30/30/30    & 0    & 0    & 0    & 0    & 0    & 0     \\
			5            & 0     & 0    & 0    & 30/30/30    & 0    & 0    & 0    & 0    & 0     \\
			6            & 0     & 0    & 0    & 0    & 30/30/30    & 0    & 0    & 0    & 0     \\
			7            & 0     & 0    & 0    & 0    & 0    & 30/30/30    & 0    & 0    & 0     \\
			8            & 0     & 0    & 0    & 0    & 0    & 0    & 30/30/30    & 0    & 0     \\
			9            & 0     & 0    & 0    & 0    & 0    & 0    & 0    & 30/30/30    & 0     \\
			10           & 0     & 0    & 0    & 0    & 0    & 0    & 0    & 0    & 30/30/30     \\ \bottomrule
		\end{tabular}%
	}
\end{table}


\begin{table}[]
	\small
	\renewcommand{\arraystretch}{1.5}%
	\centering
	\caption[Estimated dimensionalities for $n$~=~5000 spectra, SNR~=~100:1]{Comparison of estimated dimensionality for $n$~=~5000 spectra, SNR~=~100:1}
	\label{tab:ItableS5}
	\begin{tabular}{@{}lp{1.3cm}p{1.3cm}p{1.3cm}p{1.3cm}p{1.3cm}p{1.3cm}p{1.3cm}@{}}
		\toprule
		Ground truth & CD$^{[a]}$    & HFC$^{[b]}$   & AIC   & MDL   & FIF   & $\lambda_{95}$   & $\lambda_{99}$       \\ \midrule
		2 & 2 & 2 & 2 & 2 & 2 & 1 & 1 \\
		3 & 3 & 3 & 3 & 3 & 3 & 2 & 2 \\
		4 & 4 & 4 & 4 & 4 & 4 & 3 & 3 \\
		5 & 5 & 4 & 5 & 5 & 5 & 4 & 4 \\
		6 & 6 & 6 & 6 & 6 & 6 & 4 & 5 \\
		7 & 7 & 6 & 7 & 7 & 7 & 6 & 6 \\
		8 & 8 & 5 & 8 & 8 & 8 & 6 & 7 \\
		9 & 9 & 7 & 9 & 9 & 9 & 7 & 8 \\
		10 & 10 & 6 & 10 & 10 & 10 & 8 & 9 \\ \bottomrule
		\multicolumn{8}{p{11.7cm}}{\footnotesize \begin{tabular}[c]{@{}p{\linewidth}@{}}$^{[a]}$ Based on 30 repeats of the Chemical Dimensionality algorithm. Confusion matrix in Table \ref{tab:ItableS9}.  \\ $^{[b]}$ False alarm probability of $10^{-5}$.  \end{tabular}}     
	\end{tabular}
\end{table}

\begin{table}[h!]
	\small
	\renewcommand{\arraystretch}{1.5}%
	\centering
	\caption[Estimated dimensionalities for $n$~=~5000 spectra, SNR~=~50:1]{Comparison of estimated dimensionality for $n$~=~5000 spectra, SNR~=~50:1}
	\label{tab:ItableS6}
	\begin{tabular}{@{}lp{1.3cm}p{1.3cm}p{1.3cm}p{1.3cm}p{1.3cm}p{1.3cm}p{1.3cm}@{}}
		\toprule
		Ground truth & CD$^{[a]}$    & HFC$^{[b]}$   & AIC   & MDL   & FIF   & $\lambda_{95}$   & $\lambda_{99}$       \\ \midrule
		2 & 2 & 2 & 2 & 2 & 2 & 1 & 1 \\
		3 & 3 & 3 & 3 & 3 & 3 & 2 & 2 \\
		4 & 4 & 4 & 4 & 4 & 4 & 3 & 3 \\
		5 & 5 & 4 & 5 & 5 & 5 & 4 & 4 \\
		6 & 6 & 6 & 6 & 6 & 6 & 4 & 5 \\
		7 & 7 & 6 & 7 & 7 & 7 & 6 & 6 \\
		8 & 8 & 5 & 8 & 8 & 8 & 6 & 7 \\
		9 & 9 & 7 & 9 & 9 & 9 & 7 & 8 \\
		10 & 10 & 6 & 10 & 10 & 10 & 8 & 9 \\ \bottomrule
		\multicolumn{8}{p{11.7cm}}{\footnotesize \begin{tabular}[c]{@{}p{\linewidth}@{}}$^{[a]}$ Based on 30 repeats of the Chemical Dimensionality algorithm. Confusion matrix in Table \ref{tab:ItableS10}.  \\ $^{[b]}$ False alarm probability of $10^{-5}$.  \end{tabular}}     
	\end{tabular}
\end{table}

\begin{table}[h!]
	\small
	\renewcommand{\arraystretch}{1.5}%
	\centering
	\caption[Estimated dimensionalities for $n$~=~5000 spectra, SNR~=~20:1]{Comparison of estimated dimensionality for $n$~=~5000 spectra, SNR~=~20:1}
	\label{tab:ItableS7}
	\begin{tabular}{@{}lp{1.3cm}p{1.3cm}p{1.3cm}p{1.3cm}p{1.3cm}p{1.3cm}p{1.3cm}@{}}
		\toprule
		Ground truth & CD$^{[a]}$    & HFC$^{[b]}$   & AIC   & MDL   & FIF   & $\lambda_{95}$   & $\lambda_{99}$       \\ \midrule
		2 & 2 & 2 & 2 & 2 & 2 & 1 & 164 \\
		3 & 2(\SI{83}{\percent}) & 3 & 3 & 3 & 3 & 2 & 29 \\
		4 & 4 & 4 & 4 & 4 & 4 & 3 & 285 \\
		5 & 5 & 4 & 5 & 5 & 5 & 4 & 209 \\
		6 & 6 & 6 & 6 & 6 & 6 & 5 & 256 \\
		7 & 7 & 6 & 7 & 7 & 7 & 6 & 297 \\
		8 & 8 & 5 & 8 & 8 & 8 & 7 & 219 \\
		9 & 9(\SI{97}{\percent}) & 7 & 9 & 9 & 9 & 7 & 271 \\
		10 & 10 & 6 & 10 & 10 & 10 & 8 & 365 \\ \bottomrule
		\multicolumn{8}{p{11.7cm}}{\footnotesize \begin{tabular}[c]{@{}p{\linewidth}@{}}$^{[a]}$ Based on 30 repeats of the Chemical Dimensionality algorithm. Confusion matrix in Table \ref{tab:ItableS11}.  \\ $^{[b]}$ False alarm probability of $10^{-5}$.  \end{tabular}}     
	\end{tabular}
\end{table}

\begin{table}[h!]
	\small
	\renewcommand{\arraystretch}{1.5}%
	\centering
	\caption[Estimated dimensionalities for $n$~=~5000 spectra, SNR~=~10:1]{Comparison of estimated dimensionality for $n$~=~5000 spectra, SNR~=~10:1}
	\label{tab:ItableS8}
	\begin{tabular}{@{}lp{1.3cm}p{1.3cm}p{1.3cm}p{1.3cm}p{1.3cm}p{1.3cm}p{1.3cm}@{}}
		\toprule
		Ground truth & CD$^{[a]}$    & HFC$^{[b]}$   & AIC   & MDL   & FIF   & $\lambda_{95}$   & $\lambda_{99}$       \\ \midrule
		2  & 2 & 2 & 2 & 2 & 2 & 35 & 653  \\
		3  & 3(\SI{63}{\percent})& 3 & 3 & 3 & 3 & 2 & 580 \\
		4  & 5 & 4 & 4 & 4 & 4 & 160 & 713 \\
		5  & 5 & 4 & 5 & 5 & 5 & 81 & 675  \\
		6  & 6 & 6 & 6 & 6 & 6 & 130 & 699 \\
		7  & 7 & 6 & 7 & 7 & 7 & 171 & 718 \\
		8  & 8 & 5 & 8 & 8 & 8 & 91 & 680  \\
		9  & 8 & 7 & 9 & 9 & 9 & 145 & 705 \\
		10 & 10 & 6 & 10 & 10 & 10 & 243 & 749 \\ \bottomrule
		\multicolumn{8}{p{11.7cm}}{\footnotesize \begin{tabular}[c]{@{}p{\linewidth}@{}}$^{[a]}$ Based on 30 repeats of the Chemical Dimensionality algorithm. Confusion matrix in Table \ref{tab:ItableS12}.  \\ $^{[b]}$ False alarm probability of $10^{-5}$.  \end{tabular}}     
	\end{tabular}
\end{table}

\begin{table}[h!]
	\small
	\renewcommand{\arraystretch}{1.5}%
	\centering
	\caption[CD-algorithm confusion matrix for 5000 samples, SNR = 100:1]{Confusion matrix of CD-algorithm for 30 repeats, with endmembers mixed to $n$~=~5000 spectra, SNR~=~100:1 Chemical Dimensionality Algorithm}
	\label{tab:ItableS9}
	\begin{tabular}{@{}lp{1cm}p{1cm}p{1cm}p{1cm}p{1cm}p{1cm}p{1cm}p{1cm}p{1cm}@{}}
		\toprule
		& \multicolumn{9}{c}{Chemical Dimensionality Algorithm} \\ \cmidrule(r){2-10}
		Ground truth & 2    & 3   & 4   & 5   & 6   & 7   & 8   & 9   & 10   \\ \midrule
		2            & 30    & 0    & 0    & 0    & 0    & 0    & 0    & 0    & 0     \\
		3            & 0     & 30   & 0    & 0    & 0    & 0    & 0    & 0    & 0     \\
		4            & 0     & 0    & 30   & 0    & 0    & 0    & 0    & 0    & 0     \\
		5            & 0     & 0    & 0    & 30   & 0    & 0    & 0    & 0    & 0     \\
		6            & 0     & 0    & 0    & 0    & 30   & 0    & 0    & 0    & 0     \\
		7            & 0     & 0    & 0    & 0    & 0    & 30   & 0    & 0    & 0     \\
		8            & 0     & 0    & 0    & 0    & 0    & 0    & 30   & 0    & 0     \\
		9            & 0     & 0    & 0    & 0    & 0    & 0    & 0    & 30   & 0     \\
		10           & 0     & 0    & 0    & 0    & 0    & 0    & 0    & 0    & 30     \\ \bottomrule
	\end{tabular}
\end{table}

\begin{table}[h!]
	\small
	\renewcommand{\arraystretch}{1.5}%
	\centering
	\caption[CD-algorithm confusion matrix for 5000 samples, SNR = 50:1]{Confusion matrix of CD-algorithm for 30 repeats, with endmembers mixed to $n$~=~5000 spectra, SNR~=~50:1 Chemical Dimensionality Algorithm}
	\label{tab:ItableS10}
	\begin{tabular}{@{}lp{1cm}p{1cm}p{1cm}p{1cm}p{1cm}p{1cm}p{1cm}p{1cm}p{1cm}@{}}
		\toprule
		& \multicolumn{9}{c}{Chemical Dimensionality Algorithm} \\ \cmidrule(r){2-10}
		Ground truth & 2    & 3   & 4   & 5   & 6   & 7   & 8   & 9   & 10   \\ \midrule
		2            & 30    & 0    & 0    & 0    & 0    & 0    & 0    & 0    & 0     \\
		3            & 0     & 30   & 0    & 0    & 0    & 0    & 0    & 0    & 0     \\
		4            & 0     & 0    & 30   & 0    & 0    & 0    & 0    & 0    & 0     \\
		5            & 0     & 0    & 0    & 30   & 0    & 0    & 0    & 0    & 0     \\
		6            & 0     & 0    & 0    & 0    & 30   & 0    & 0    & 0    & 0     \\
		7            & 0     & 0    & 0    & 0    & 0    & 30   & 0    & 0    & 0     \\
		8            & 0     & 0    & 0    & 0    & 0    & 0    & 30   & 0    & 0     \\
		9            & 0     & 0    & 0    & 0    & 0    & 0    & 0    & 30   & 0     \\
		10           & 0     & 0    & 0    & 0    & 0    & 0    & 0    & 0    & 30     \\ \bottomrule
	\end{tabular}
\end{table}

\begin{table}[h!]
	\small
	\renewcommand{\arraystretch}{1.5}%
	\centering
	\caption[CD-algorithm confusion matrix for 5000 samples, SNR = 20:1]{Confusion matrix of CD-algorithm for 30 repeats, with endmembers mixed to $n$~=~5000 spectra, SNR~=~20:1 Chemical Dimensionality Algorithm}
	\label{tab:ItableS11}
	\begin{tabular}{@{}lp{1cm}p{1cm}p{1cm}p{1cm}p{1cm}p{1cm}p{1cm}p{1cm}p{1cm}@{}}
		\toprule
		& \multicolumn{9}{c}{Chemical Dimensionality Algorithm} \\ \cmidrule(r){2-10}
		Ground truth & 2    & 3   & 4   & 5   & 6   & 7   & 8   & 9   & 10   \\ \midrule
		2            & 30    & 0    & 0    & 0    & 0    & 0    & 0    & 0    & 0     \\
		3            & 25    & 3    & 2    & 0    & 0    & 0    & 0    & 0    & 0     \\
		4            & 0     & 0    & 30   & 0    & 0    & 0    & 0    & 0    & 0     \\
		5            & 0     & 0    & 0    & 30   & 0    & 0    & 0    & 0    & 0     \\
		6            & 0     & 0    & 0    & 0    & 30   & 0    & 0    & 0    & 0     \\
		7            & 0     & 0    & 0    & 0    & 0    & 30   & 0    & 0    & 0     \\
		8            & 0     & 0    & 0    & 0    & 0    & 0    & 30   & 0    & 0     \\
		9            & 0     & 0    & 0    & 0    & 0    & 0    & 1    & 29   & 0     \\
		10           & 0     & 0    & 0    & 0    & 0    & 0    & 0    & 0    & 30     \\ \bottomrule
	\end{tabular}
\end{table}

\begin{table}[h!]
	\small
	\renewcommand{\arraystretch}{1.5}%
	\centering
	\caption[CD-algorithm confusion matrix for 5000 samples, SNR = 10:1]{Confusion matrix of CD-algorithm for 30 repeats, with endmembers mixed to $n$~=~5000 spectra, SNR~=~10:1 Chemical Dimensionality Algorithm}
	\label{tab:ItableS12}
	\begin{tabular}{@{}lp{1cm}p{1cm}p{1cm}p{1cm}p{1cm}p{1cm}p{1cm}p{1cm}p{1cm}@{}}
		\toprule
		& \multicolumn{9}{c}{Chemical Dimensionality Algorithm} \\ \cmidrule(r){2-10}
		Ground truth & 2    & 3   & 4   & 5   & 6   & 7   & 8   & 9   & 10   \\ \midrule
		2            & 30    & 0    & 0    & 0    & 0    & 0    & 0    & 0    & 0     \\
		3            & 4     & 19   & 6    & 1    & 0    & 0    & 0    & 0    & 0     \\
		4            & 0     & 0    & 0    & 30   & 0    & 0    & 0    & 0    & 0     \\
		5            & 0     & 0    & 0    & 30   & 0    & 0    & 0    & 0    & 0     \\
		6            & 0     & 0    & 0    & 0    & 30   & 0    & 0    & 0    & 0     \\
		7            & 0     & 0    & 0    & 0    & 0    & 30   & 0    & 0    & 0     \\
		8            & 0     & 0    & 0    & 0    & 0    & 0    & 30   & 0    & 0     \\
		9            & 0     & 0    & 0    & 0    & 0    & 0    & 30   & 0    & 0     \\
		10           & 0     & 0    & 0    & 0    & 0    & 0    & 0    & 0    & 30    \\ \bottomrule
	\end{tabular}
\end{table}


\FloatBarrier
%
%
%
%
%


\end{document}